\documentclass[11pt]{article}
\usepackage[utf8]{inputenc}
\pdfoutput=1
\usepackage{amssymb,amsmath,mathrsfs,enumerate}
\usepackage{graphicx,rotate,multicol}
\usepackage{float}
\usepackage{tocloft}
\usepackage{subcaption}
\usepackage[margin=10pt,labelfont=bf]{caption}
\usepackage{cite}
\usepackage{soul}
\usepackage[colorlinks=true,
            linkcolor=blue,
            urlcolor=blue,
            citecolor=teal]{hyperref}
\usepackage{multirow}
\usepackage{placeins}
\usepackage[normalem]{ulem}
\makeatletter
\let\Hy@linktoc\Hy@linktoc@page
\makeatother
\usepackage{color}
\definecolor{ourcolor}{rgb}{0.7, 0.25, 0.05}
\usepackage{tikz,braket}
\long\def\rpl#1!!#2!!{\textcolor{red}{#1} \textcolor{blue}{#2}}

\def \order(#1){{\mathcal O} \left(#1 \right)}

\evensidemargin=\oddsidemargin
\allowdisplaybreaks

\title{\color{black}{\bf Impact of galactic distributions in celestial capture of dark matter}}

\author {\bf Debajit Bose\footnote{debajitbose550@gmail.com} \hspace{1pt} and \hspace{1pt} Sambo Sarkar\footnote{sambosarkar92@gmail.com}
	\\[10pt]
	\small\em Department of Physics, Indian Institute of Technology Kharagpur, Kharagpur 721302, India \\}

\date{}

\begin{document} 
	
	\maketitle

	\begin{abstract}

Celestial capture of dark matter provides a useful handle for constraining its particulate properties. The capture formalism is sensitive to the phase space distribution of dark matter in the vicinity of the celestial object. This article aims to systematically study the impact of uncertainties and the influence of cosmological simulations on the rate at which dark matter particles are captured inside a variety of celestial objects. Going beyond the framework of the Maxwell-Boltzmann distribution or the standard halo model, we take up pragmatic dark matter velocity distributions motivated by observations or cosmological simulations. Within the limits of the standard halo model, we report a maximum $\sim 20\%$ change in the capture rate. This number can go up to $\sim 200\%$ if dark matter particles within the galactic halo are favored to have an empirical velocity distribution profile when well-resolved and sophisticated cosmological simulations are employed to extract their parametric values.
	\end{abstract}

\section{Introduction}
\label{sec:intro}

Astrophysical and cosmological explorations have presented us with ample evidence, advocating the predominance of dark matter (DM) out of the total matter content of the universe \cite{Clowe:2006eq,Planck:2018vyg}. A phenomenologically interesting conjecture would be to consider particulate DM,  having non-gravitational interactions with standard model (SM) particles \cite{Goodman:1984dc,Bertone:2010zza}. A handful of terrestrial detectors which aim at detecting DM and probing such interactions, are yet to report a conclusive DM detection signature. \cite{Cui:2017nnn,XENON:2018voc}. In addition to these ground-based direct detection experiments, there has been a surge to probe DM-SM interactions using astrophysical observables and cosmological techniques \cite{Iocco:2008xb,Bramante:2021dyx}. Among such probes, which are collectively known in the literature as indirect search channels, the capture of DM within celestial bodies has been argued to be a promising blueprint in the search for particulate DM \cite{Slatyer:2017sev,Leane:2020liq,Boddy:2022knd}.

As astrophysical bodies move within a halo, DM particles tend to get trapped inside the gravitational potential of the astrophysical object. On its entry into the stellar interior, DM scatters from the constituents of the celestial object. Due to its non-gravitational interactions with SM particles, it loses a sufficient fraction of its energy. Single or multiple scattering events can reduce the initial energy of DM particles below the escape energy of the concerned celestial object \cite{Press1985CaptureBT,Gould:1987ir}, thereby getting captured in the process. The accumulated DM particles can annihilate to SM states which can give rise to heating signatures of neutron stars \cite{Kouvaris:2007ay,Bramante:2017xlb,Raj:2017wrv,Bell:2018pkk,Acevedo:2019agu,Joglekar:2019vzy,Bell:2020jou,Joglekar:2020liw,Bell:2020lmm,Ilie:2020vec,Maity:2021fxw,Ilie:2021iyh,Anzuini:2021lnv,McKeen:2021jbh,Coffey:2022eav}, white dwarfs \cite{McCullough:2010ai,Dasgupta:2019juq,Bell:2021fye} or exoplanets \cite{Leane:2020wob,Bramante:2022pmn}. In a contrasting scenario, the final state SM products or long-lived mediators of DM annihilation can lead to an observable neutrino or gamma-ray flux, respectively, at terrestrial detectors \cite{PhysRevLett.55.257,IceCube:2012ugg,Pospelov:2008jd,Batell:2009zp,Leane:2017vag,Bell:2021pyy,Leane:2021ihh,Bose:2021yhz,Bose:2021cou}. In some extreme cases, the accumulation of DM particles inside a celestial object can lead to the formation of black holes \cite{PhysRevD.40.3221,GOULD1990337,Bertone:2007ae,Bell:2013xk,Garani:2018kkd,Acevedo:2020gro}, ignition of supernova \cite{Bramante:2015cua,Janish:2019nkk}, alteration of periods in binary stars \cite{Hassani:2020uhd}, etc.

A significant quantity in the study of DM capture is the rate at which DM particles get trapped inside the stars.  The uncertainties associated with the capture rate arise from the  DM density and the velocity distribution of DM particles in the neighbourhood of their celestial bodies.  The dependence on the former is straightforward as the capture rate scales linearly with local DM density.  The majority of results derived in the literature are presented based on the Maxwell-Boltzmann (MB) velocity profile \cite{Press1985CaptureBT,Gould:1987ir,Kouvaris:2007ay,Bramante:2017xlb,Bell:2020jou,Dasgupta:2019juq,Garani:2017jcj,Busoni:2017mhe}. In this article, we explore the uncertainties in the capture rate primarily from observationally motivated, non-SHM velocity distributions of DM and other governing parameters like DM velocity dispersion and the galactic halo escape velocity.  The authors of \cite{Nunez-Castineyra:2019odi,Lopes:2020dau} has considered the uncertainties related to DM capture inside the sun and neutron stars. We systematically study and catalog uncertainties and effects on DM capture for different DM capture scenarios for the first time. However, in this work, we have not considered the uncertainties arising due to internal constitution, equation of state, mass, radial extent, and atmospheric modeling of individual celestial bodies. The uncertainties related to different equations of state, the velocity dispersion of celestial bodies, and the detectability of dark heating for neutron stars have been studied in \cite{Chatterjee:2022dhp}. The percentage variations due to the mentioned parameters require separate attention and remain outside this work's mandate. However, we have provided approximate estimates of the uncertainties arising from variations in mass and radius, wherever possible.

The rest of the paper is organized as follows.  In section \ref{sec:formalism}, we describe the general formalism associated with the capture of DM inside celestial objects. In section \ref{sec:velocity_dis}, we discuss the origin of the astrophysical parameters and their associated errors, which lead to variations in the DM capture rate.  In section \ref{sec:simu}, we focus on the effects of cosmological simulations on certain empirical velocity distributions.  In sections \ref{sec:cap_all} and \ref{sec:cap_sun}, we catalog the effective change in the capture rate from the standard choices, for five different  celestial objects. In section \ref{sec:comp} we provide a systematic comparison between the results obtained from astrophysical data to the ones obtained using cosmological simulations. Finally we conclude the findings of our work in section \ref{sec:conclude}.

\section{General formalism of DM capture}
\label{sec:formalism}

In this section, we review the basic formalism of DM capture in astrophysical objects. As a celestial body moves in the galactic halo, the DM particles in the halo get focused on the body due to its steep gravitational potential. The velocity of the DM particle upon reaching the surface of the celestial object is given by $w = \sqrt{u_{\chi}^2+v_{\rm esc}^2}$, where $v_{\rm esc}$ is the escape velocity of the concerned object and $u_{\chi}$ is the velocity of the DM in the halo. These DM particles scatter with the SM constituents of the celestial body, and if, after collision, the DM particle attains a velocity that is less than $v_{\rm esc}$, then it gets captured within the celestial body. The rate of DM capture in an astronomical object is given by \cite{Dasgupta:2019juq,Bramante:2017xlb,Ilie:2020vec},
\begin{equation}
\label{eq:caprate}
C_{\rm tot} = \sum_{\rm S} C_{\rm S} = \sum_{\rm S} \pi \, R^2 \, p_{\rm S}(\tau) \, \left( \frac{\rho_{\chi}}{m_{\chi}} \right) \int_{0}^{u_{\rm esc}} \dfrac{f(u_{\chi}) \, du_{\chi}}{u_{\chi}} \, \left( u_{\chi}^2+v_{\rm esc}^2 \right) \, g_{\rm S}(u_{\chi}),
\end{equation}
where $m_{\chi}$, $\rho_{\chi}$ and $f(u_{\chi})$ are the mass, DM density at the location of the celestial object and the velocity distribution profile of DM in the galactic rest frame. $\tau$ is the optical depth defined as $\tau = 3\sigma {\cal N}_t/(2\pi R^2)$, ${\cal N}_t$ being the total number of targets in the astrophysical object and $\sigma$ is the DM-nuclei interaction cross-section and $g_{\rm S}$ denotes the probability that the final velocity of DM after $\rm S$ scatterings becomes less than $v_{\rm esc}$ \cite{Dasgupta:2019juq}. The probability that DM particle scatters S times within the celestial object is given by,
\begin{equation}
\label{prob}
p_{\rm S}(\tau)=2\int_{0}^{1}dy\,\dfrac{ye^{-y\tau}(y\tau)^S}{S!}
\end{equation}
Evidently, from Eq. \eqref{eq:caprate}, the uncertainty in the capture rate depends on the DM density and the velocity distribution profile. The uncertainty due to DM density is trivial as it increases linearly when we move toward denser patches of the universe. In this paper, we concentrate on the uncertainties arising due to the different velocity distribution profiles $f(u_{\chi})$, incorporating the velocity dispersion and galactic escape velocity from both observations and simulations while fixing the DM density.

\section{Uncertainties associated with the capture process}
\label{sec:velocity_dis}
As evident from Eq. \eqref{eq:caprate}, the uncertainties in estimating the rate at which DM gets trapped within celestial objects depend primarily on the local DM density, local DM velocity distribution, galactic escape velocity, and DM velocity dispersion.  In the following subsection, we will discuss the effects of these parameters, which are highly dependent on the achievements made in astrophysical measurements.

\begin{enumerate}

\item \textbf{DM density:}

From Eq. \eqref{eq:caprate}, we see that any change in the local DM density $\rho_{\chi}$, would lead to a proportional shift in the capture rate, independent of the celestial object in consideration. However, recent works show that DM can also be captured within a distribution of stars \cite{Leane:2021ihh,Bose:2021yhz}. In such a scenario, the total capture rate due to a collection of stars does not depend linearly on the DM density. Throughout this work, we will concentrate on DM capture by a particular celestial body and fix the value of DM density to be the local DM density at $0.4~ \rm GeV\, cm^{-3}$. As we are concerned with the relative difference in the capture rate, any change owing to $\rho_{\chi}$ will not give any interesting phenomenological signatures.

\item \textbf{Circular velocity of the Sun:}

The circular velocity of the Sun with respect to the galactic center is considered to be the velocity dispersion ($u_0$) of the DM velocity distribution. Orbit of the GD-1 stellar stream \cite{Koposov:2009hn} had constrained $u_0$ in the range $221 \pm 18 ~\rm km/s$. Similar ranges exist between $225 \pm 29 ~\rm km/s$ in consonance to the kinematics of maser \cite{McMillan:2009yr}. These measurements give around $10\%$ error in  $u_0$. A more precise assessment of $u_0$ can be done using the measurement of the apparent motion of Sgr A$^*$, relative to a distant quasar \cite{Reid:2004rd, Bland_Hawthorn_2016}, fixing the total angular velocity of the Sun ($(u_0+V_{\circledast})/R_{\circledast}$) within $30.24 \pm 0.12 ~\rm km \,s^{-1} kpc^{-1}$. Also the GRAVITY collaboration has recently estimated the value of $R_{\circledast}$ with a considerable high accuracy at $8.122 \pm 0.031$ kpc \cite{Abuter:2018drb}. From the results of these two observations, the circular velocity of the Sun has been estimated to be around $233 \pm 3\, \rm km/s$ \cite{Evans:2018bqy}. We will consider this latest empirical estimation as the benchmark choice for $u_0$ in this article. Other relevant observations also lead to similar results \cite{Gillessen:2009ht, Reid_2014, Eilers_2019, Hogg_2019}. We will explore the impact of the deviations of this value within the uncertainty limits on the capture rate of DM. $u_0$ being related to the standard deviation of the velocity distribution, any increment in $v_0$  would flatten the distribution. One would thereby expect an increase of DM particles at the high velocity tail of the distribution, making it difficult for these high velocity particles to get captured by the gravitational potential of the star. The effect will be reversed for a decrement in $u_0$.

\item \textbf{Galactic escape velocity:}

Escape velocity of a galactic halo is the velocity of a particle below which it does not remain bound to the gravitational potential the galaxy. Measurements from the high velocity stars in the RAVE survey determines $u_{\rm esc}$ in the range $498-608 $ km/s \cite{Smith:2006ym} having a median of $544$ km/s. Recent estimates from the velocities of $2850$ halo stars from the \textit{Gaia} velocity survey Data Release-2 \cite{Monari_2018}, has been revised the local escape speed within $580 \pm 63$ km/s. With a prior estimation from simulations, and a more localized sample of $2300$ high velocity counter-rotating stars, the escape speed has been obtained to be $528^{+24}_{-25}$ km/s in \cite{Deason_2019}. This is also in consonance with the previous results. We will be using the central value of the latter as the benchmark escape velocity of DM particles of a MW like halo. For a particular velocity distribution, an increase in the $u_{\rm esc}$ keeping the $u_0$ fixed will lead to an increase of DM particles in the high velocity tail, suppressing the capture probability which is largely dominated by low velocity DM particles.

\item \textbf{DM velocity distribution}:

Equation \eqref{eq:caprate} thoroughly encapsulates the impact of DM velocity distribution, velocity dispersion and the escape velocity on the capture rate. DM particles in MW like galaxies are conventionally assumed to have a Maxwell-Boltzmann velocity distribution, with a sharp cut-off at the galactic escape velocity \cite{LyndenBell:1966bi,1990ApJ...353..486L}. This description is popularly known as the standard halo model (SHM) given by,
\begin{equation}
f(\mathbf{u_{\chi}})=
\begin{cases}
\frac{1}{N}\left[\exp\left(-\frac{|\mathbf{u_{\chi}}|^{2}}{u_0^2}\right)\right] &  |\mathbf{u_{\chi}}| \leq u_{ \rm esc} \\
0 &  |\mathbf{u_{\chi}}| > u_{\rm esc}.
\end{cases}
\label{eq:MB}
\end{equation}
Throughout  this work we will consider the benchmark choice of the two SHM parameters as $u_{0} = 233 \, \rm km/s $ and $u_{\rm esc} = 528 \, \rm km/s $, motivated from the above discussions. 
However the SHM provides an oversimplified description of the speed distribution which might not capture in its entirety the genuine distributions of low and high velocity particles in the MW halo. 

\end{enumerate}

\section{Role of Cosmological simulations}
\label{sec:simu}

Cosmological simulations are widely employed to generate selected patches of the universe, primarily aiming to understand the dynamics of our immediate neighborhood. These include the Milky Way (MW), its DM halo, bulge component, and other substructures. Baryon physics and anisotropic effects are often incorporated using hydrodynamic techniques.

\subsection{Cosmological simulations}
\label{subsec:simunames}

 A significant motivation for this work has been derived from some of the well resolved cosmological simulations like NIHAO \cite{Wang:2015jpa}, APOSTLE \cite{Sawala:2015cdf},  AURIGA\cite{Grand:2016mgo},  IllustrisTNG \cite{Pillepich:2017jle}, FIRE\cite{Hopkins:2017ycn}, ARTEMIS \cite{Font_2020}, etc.  Generating initial conditions from cosmological perturbation theory and incorporating the effects of hierarchical structure formation, mergers, etc, these simulations indicate that the SHM may not give an accurate description of a Milky Way-like halo. Such simulations prescribe the actual DM velocity distribution in haloes to depart from a Maxwellian distribution \cite{Vogelsberger:2008qb,Kuhlen:2009vh,Lisanti:2010qx}. Due to its sharp cut-off at the galactic escape velocity, the SHM tends to over predict the number of high energetic DM particles available. This development has driven modifications in the SHM framework, in order to reconcile the discrepancies  between the astrophysical observations and the aforementioned cosmological simulations \cite{Kavanagh:2013eya, Lee:2014cpa,Kelso:2016qqj, Sloane:2016kyi}. A discussion on the non-SHM, isotropic velocity distributions are now in order. 

\subsection{Empirical distributions from simulations}
\label{sec:simuncert}

One of the empirically motivated models is the King velocity distribution, which determines the maximum velocity cut-off in a self-consistent manner \cite{2008gady.book.....B}. The Double Power Law (DPL) has been argued to explain with much precision, the high-velocity dependence of double power density profiles like that of the NFW distribution \cite{Lisanti:2010qx}. The Tsallis model is also a theoretical distribution based on the Gibbs entropy and inspired by the Tsallis statistics \cite{Tsallis:1987eu}. Another interesting velocity distribution, the Mao \emph{et al.} hypothesized in \cite{Evans:2018bqy}, exhibits a strong correlation of particle velocities to their position and the characteristic radius of the simulated haloes. In this work, we have only considered the impact of isotropic velocity distributions.
 
\begin{enumerate}

\item \textbf {King distribution:}
\label{king}

Taking into account the finite extent of a DM halo, the King distribution \cite{King:1966fn} can be sketched in a self-consistent manner. Instead of the galactic escape velocity, the maximum DM particle velocity of the system $ u_{\rm max}$ determines the velocity cut-off criterion ($ u_{\rm max} < u_{\rm esc}$) for this distribution \cite{Chaudhury:2010hj}. This boundary is often called the truncated radius, which represents the physical size of a halo. Such a finite-size halo provides a more realistic description of galaxies as compared to the isothermal sphere. The distribution function can be written as
\begin{equation}
    f(\mathbf{u_{\chi}})=
    \begin{cases}
      \frac{1}{N}\left[\exp{\left(\frac{u_{\rm esc}^{2}-|\mathbf{u_{\chi}}|^{2}}{u_{0}^{2}}\right)-1}\right] &  |\mathbf{u_{\chi}}| \leq u_{\rm max} \\
      0 &  |\mathbf{u_{\chi}}| > u_{\rm max},
    \end{cases}
    \label{eq:king}
\end{equation}

\item \textbf {Double Power Law distribution:}
\label{DPL}

The Double Power Law (DPL) is an empirical velocity distribution of DM particles in the galactic halo, which has been argued to very well describe empirical matter distributions such as the NFW, Hernquist, etc \cite{Lisanti:2010qx}. The velocity distribution is expressed by the form
\begin{equation}
    f(\mathbf{u_{\chi}})=
    \begin{cases}
      \frac{1}{N}\left[\exp{\left(\frac{u_{\rm esc}^{2}-|\mathbf{u_{\chi}}|^{2}}{k u_{0}^{2}}\right)}-1\right]^{k} &  |\mathbf{u_{\chi}}| \leq u_{\rm esc} \\
      0 &  |\mathbf{u_{\chi}}| > u_{\rm esc},
    \end{cases}
\label{eq:DPLfv}
  \end{equation}  
where the symbols have their usual meaning. Taking into account the quasi-static equilibrium nature of the virialised objects \cite {LyndenBell:1966bi}, structure formation history, smooth accretion and violent relaxations \cite{Wang:2008un} accurately into cosmological simulations, favour such distributions \cite{Lisanti:2010qx}. The DPL velocity distribution smoothly goes to zero at the escape velocity in contrast to the SHM and predicts lesser number of DM particles near the tail of the velocity distribution. As evident from Eq. \ref{eq:DPLfv}, as $ k \rightarrow 0 $ the DPL distribution reduces to the SHM and for $k=1$ it tends to the King distribution.

\item \textbf {Tsallis distribution:}
\label{Tsallis}

The Tsallis velocity distribution is derived through a factorization approximation of the Tsallis statistics \cite{Tsallis:1987eu}, an abstraction of the Boltzmann-Gibbs entropy. It is extensively applied in high energy collisions \cite{Cleymans:2015lxa}, Bose-Einstein condensation \cite{MILLER2006357}, black-body radiation, early universe cosmology \cite{szczniak2018nonparametric} and  superconductivity \cite{Parvan:2019hqf}. The velocity distribution function is given by
\begin{equation}
    f(\mathbf{u_{\chi}})=
    \begin{cases}
      \frac{1}{N}\left[1-\left(1-q\right)\frac{|\mathbf{u_{\chi}}|^{2}}{u_{0}^{2}}\right]^{\frac{1}{1-q}} &  |\mathbf{u_{\chi}}| \leq u_{\rm esc} \\
      0 &  |\mathbf{u_{\chi}}| > u_{\rm esc},
    \end{cases}
    \label{eq:Tsallis}
  \end{equation}
where the symbols have their usual meaning. For $q<1$ the escape velocity is obtained by the relation $u_{\rm esc}^2=u_0^2/(1-q)$.  This built-in criterion makes the distribution more robust  compared to the SHM. However, for $q>1$, the escape velocity becomes somewhat arbitrary. In $q \to 1$ limit, the Tsallis distribution goes to the Gaussian form of the SHM. Further from Eq. \eqref{eq:Tsallis} it is evident that this distribution predicts a continuous and smooth fall near the tail \cite{Vogelsberger:2008qb, Kuhlen:2009vh, Ling:2009eh}. It has been argued in  \cite{Ling:2009eh} that the  Tsallis distribution provides a better to the Milky Way like halo simulations that includes detailed baryonic physics.

\item \textbf {Mao \emph{et al.} distribution:}
\label{Mao}

In Mao \emph{et al.} \cite{Mao:2012hf}, an empirical profile having a wider peak and a steeper tail with respect to the SHM has been hypothesized as the velocity distribution of DM particles, valid over a wide shape and size of galactic haloes. The distribution is given by
\begin{equation}
    f(\mathbf{u_{\chi}})=
    \begin{cases}
      \frac{1}{N}\left[\left( u_{\rm esc}^{2}-|\mathbf{u_{\chi}}|^{2} \right)^{p} e^{-\frac{u_{\chi}}{u_{0}}}  \right] &  |\mathbf{u_{\chi}}| \leq u_{\rm esc} \\
      0 &  |\mathbf{u_{\chi}}| > u_{\rm esc},
    \end{cases}
    \label{eq:Mao}
  \end{equation}
where the symbols have their usual meaning. Some recent well resolved simulations which have considered the sequence of mergers, violent-relaxations and accretion also prefer such a velocity distribution function of its DM particles \cite{Hansen:2005yj}. This model unlike other variants of the MB distribution is not based on a Gaussian, instead it is based on an exponential distribution function with a power law cut-off at the binding energy or equivalently the escape velocity of the DM halo.  

\end{enumerate}

The values of $u_0$, $u_{\rm esc}$ and other relevant model parameters can also be derived from the DM phase space distributions obtained from some of the well resolved cosmological simulations of MW like haloes. In this work we have employed the results of some recent cosmological simulations like APOSTLE  and ARTEMIS that include baryons along-with DM. For this purpose we have used the data provided in \cite{Maity:2020wic}. Additionally, we use the DM velocity distribution of \texttt{m12f} halo from FIRE-2 simulation suite \cite{Necib:2018igl} using standard python regression tools in \texttt{Scipy} \cite{Virtanen:2019joe} and \texttt{Mathematica}, following the method discussed in \cite{Maity:2020wic}. The corresponding values for all the simulations is tabulated in Table \ref{tab:bestfit}. We provide a brief description of the simulations in appendix \ref{sec:app1}. 
\begin{table}[t]
	\centering
	\bigskip  
	\begin{tabular}{|c|c|c|c|c|c|c|c|c|c|}
		\hline
		\multirow{2}*{Simulation} & $u_{\rm esc}$ & MB & King 	& \multicolumn{2}{|c|}{DPL} & \multicolumn{2}{|c|}{ Mao } & \multicolumn{2}{|c|}{ Tsallis } \\ \cline{3-10} 
		& 		(km/s)  	& \begin{tabular}[c]{@{}c@{}}$u_0$ \\ (km/s) \end{tabular}  & \begin{tabular}[c]{@{}c@{}}$u_0$ \\ (km/s) \end{tabular} & \begin{tabular}[c]{@{}c@{}}$u_0$ \\ (km/s) \end{tabular} & $k$  & \begin{tabular}[c]{@{}c@{}}$u_0$ \\ (km/s) \end{tabular} & $p$ & \begin{tabular}[c]{@{}c@{}}$u_0$ \\ (km/s) \end{tabular} & $q$ \\ \cline{1-10} 
		
		\setlength\arraycolsep{10pt}
		
		ARTEMIS \cite{Font_2020, Poole-McKenzie:2020dbo} &  521.6 & 184.3 & 184.6 & 184.3 & 0.67& 174.7 & 3.4 & 209.4 & 0.839 \\ \hline

		APOSTLE \cite{Sawala:2015cdf, Fattahi_2016} &  646  & 224.1 & 223	 & 212.7 & 0.1 & 165   & 2.2 & 257 & 0.841 \\ 
		\hline
				
	    FIRE-2 \cite{Hopkins:2017ycn,Necib:2018igl} & 600 & 273.8 & 281 & 279 & 1.2 & 280 & 2.1 & 276 &  0.788 \\ \hline
		
	\end{tabular}
	\caption{Best fit values used to derive the capture rate for MB, King, DPL, Mao \emph{et al.}, and Tsallis  distribution profiles.}
	\label{tab:bestfit}
\end{table}

\section{Uncertainties in DM capture}
\label{sec:cap_all}

Certain unique celestial bodies existing in nature, have been explored for the capture and subsequent hunt for DM annihilation, heating or other fascinating signatures, relating to the existence of DM. An elaborate analysis of astrophysical signatures have been performed for various celestial bodies, resulting in stringent bounds on the scattering cross-section of DM and SM states. However, the limits are subject to uncertainties from DM density and velocity distributions around the celestial body. As discussed earlier, the uncertainty due to the DM density is nothing more than a scaling of the capture rate. Therefore, we can safely concentrate on the uncertainty related to the DM velocity distributions and its corresponding parametric values. In this section we make a systematic catalog for some well-motivated gravitaionally bound celestial objects. We compute the uncertainties in the capture rate due to the previously mentioned velocity profiles, using the best fit values obtained from cosmological simulations and astrophysical observations. 

The systemic effects of astrophysical uncertainties for a wide range of celestial bodies have been quantified by the relative change of capture rate from the standard MB distribution with the benchmark values using,
\begin{equation}
\Delta = \frac{ \rm C^{\rm i} - \rm C^{\rm MB}_{Bench}}{ \rm C^{\rm MB}_{Bench}}.
\label{eq:capt-effective}
\end{equation}
Here $\rm C^{\rm i}$ denotes the capture rate corresponding to the non-standard velocity distributions and/or any digression in the values of $u_0$ and $u_{\rm esc}$. In this section, we have considered the DM-nucleon cross-section to be in the order of saturation cross-section for the concerned celestial body, to emphasize the multi-scatter domain of capture framework. The relative changes in the capture rate for the optically thin and the geometric limit have been discussed and presented in appendix \ref{sec:app2}.
\subsection{Neutron stars}
\label{sec:ns}

Stars above $8\, \rm M_{\odot}$ undergo supernovae explosions at the end of their nuclear fusion cycles, leaving behind a dense core of degenerate neutrons known as neutron stars (NS). The gravitational collapse of a neutron star is balanced by the degeneracy pressure of neutrons. As the neutron stars are considered to be the densest objects in our universe apart from black holes, they are ideal to probe DM-SM interactions \cite{Bell:2020lmm}.

To study the impact of astrophysical variability on the capture rate, we consider a typical NS of mass $1.5 \, \rm M_{\odot}$ and radius $10$ km \cite{Bramante:2017xlb}. It is the core that accounts for approximately $99\%$ of the NS mass. We will consider the NS to have a neutron degenerate core of constant density and that DM majorly scatters with neutrons. We work with a DM-nucleon scattering cross-section of $10^{-45} \, \rm cm^2$.

\begin{figure}[t]
\centering
\begin{subfigure}{0.325\textwidth}
\centering
\includegraphics[width=1\linewidth]{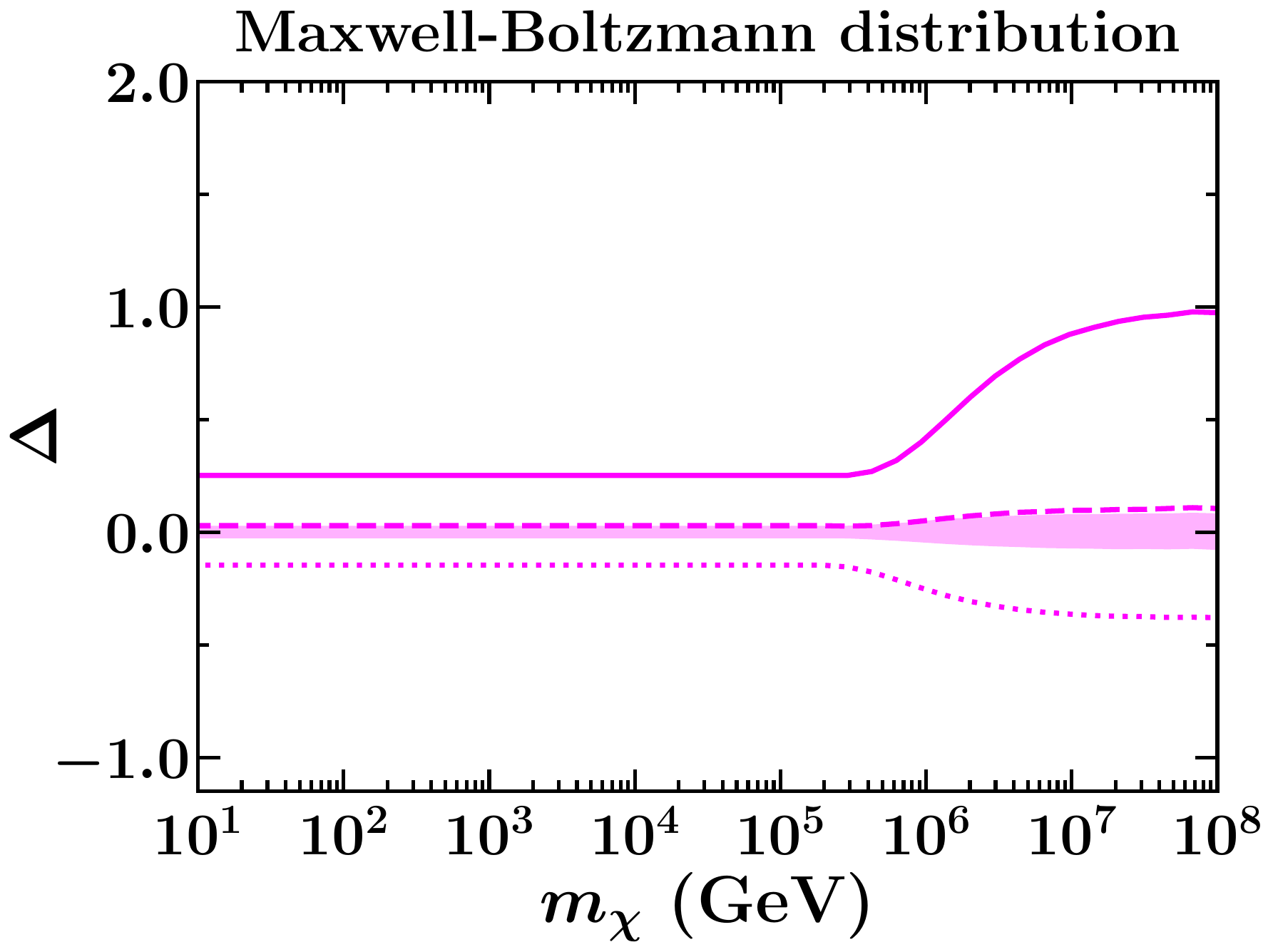} 
\caption{}
\label{sf:MB1}
\end{subfigure}
\begin{subfigure}{0.325\textwidth}
\centering
\includegraphics[width=1\linewidth]{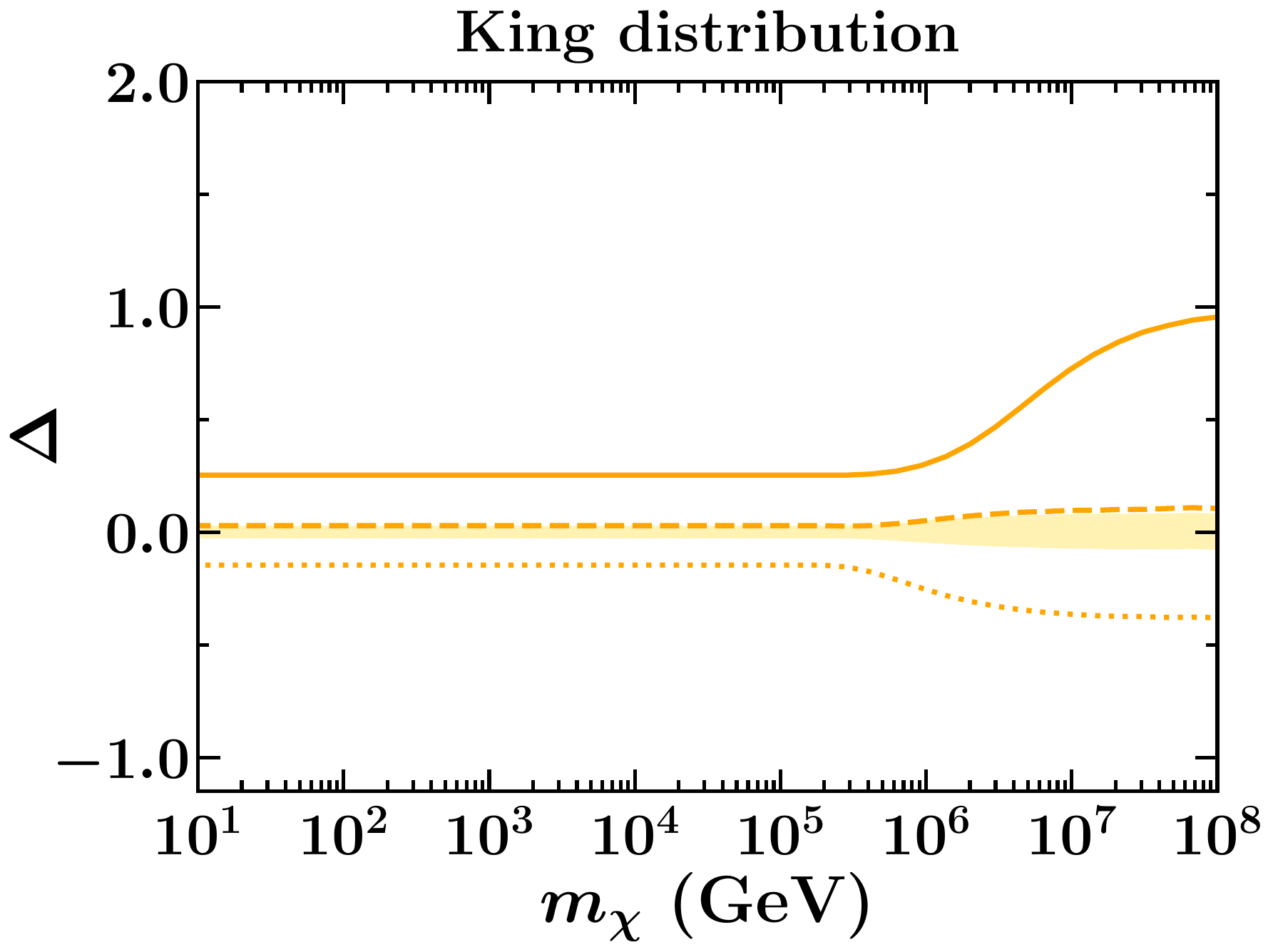} 
\caption{}
\label{sf:King1}
\end{subfigure}
\begin{subfigure}{0.325\textwidth}
\centering
\includegraphics[width=1\linewidth]{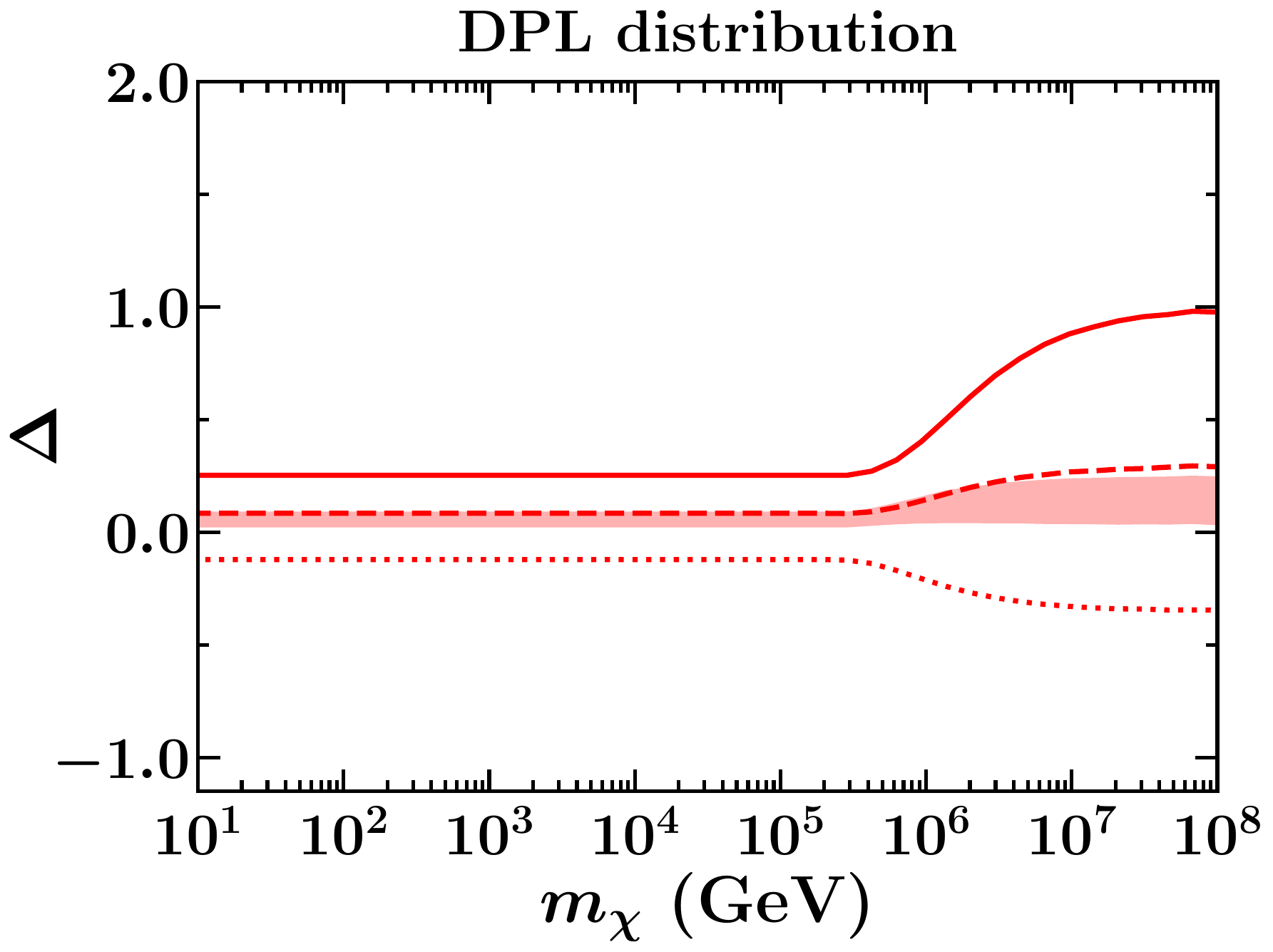} 
\caption{}
\label{sf:DPL1}
\end{subfigure}
\begin{subfigure}{0.325\textwidth}
\centering
\includegraphics[width=1\linewidth]{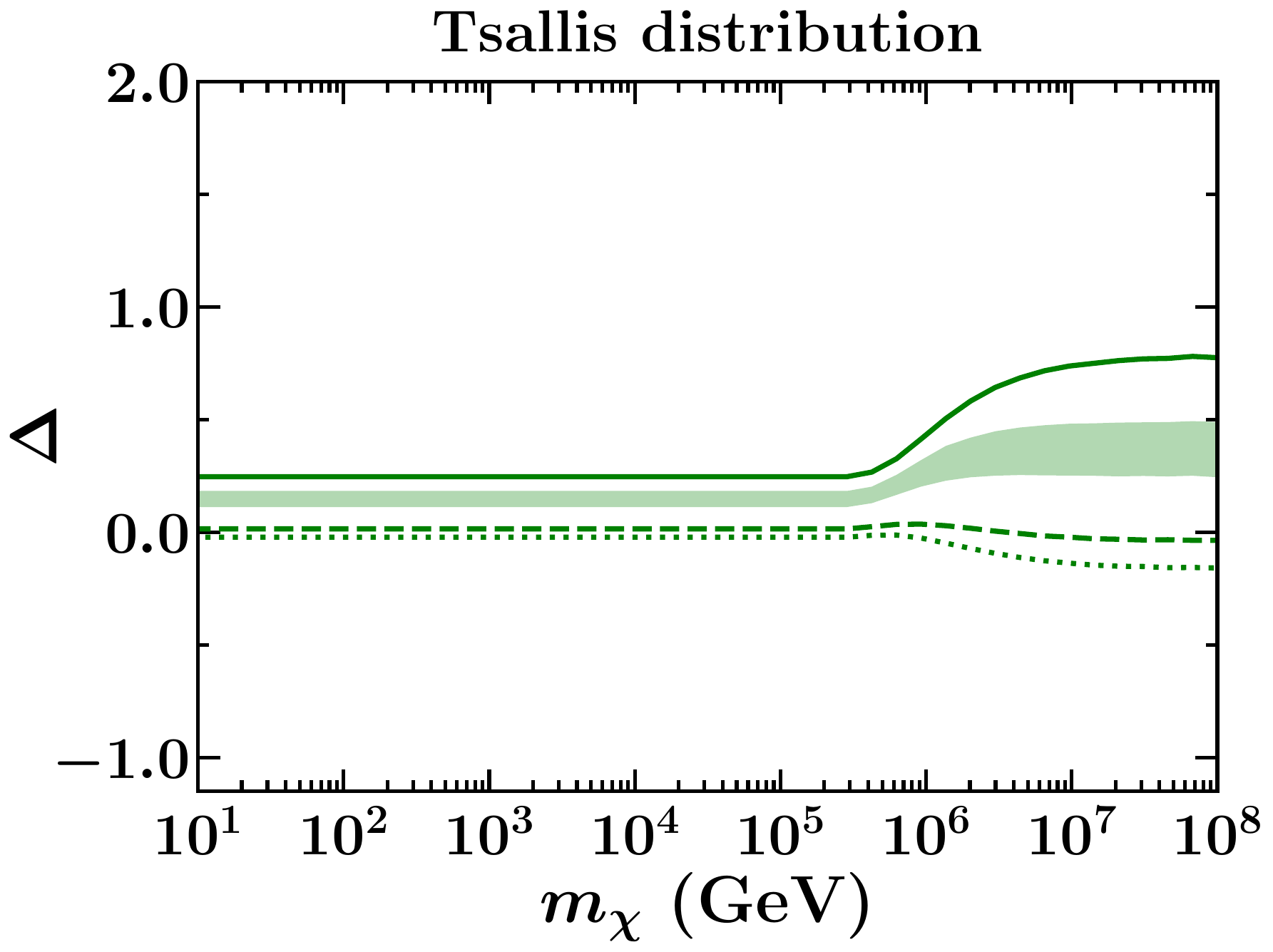} 
\caption{}
\label{sf:Tsal1}
\end{subfigure}
\begin{subfigure}{0.325\textwidth}
\centering
\hspace{0.35em} \vspace{0.55em}
\includegraphics[width=0.65\linewidth]{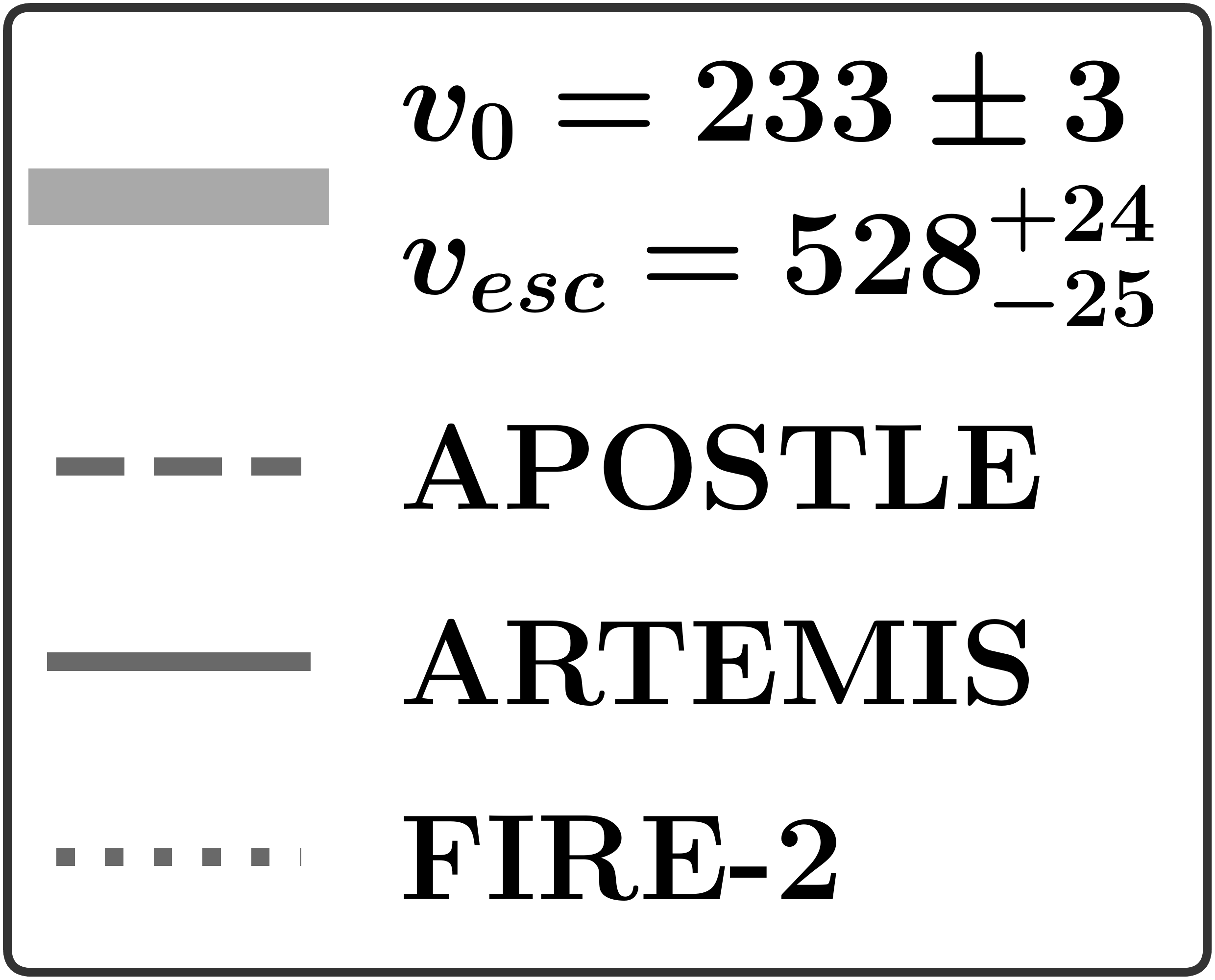} 
\caption*{}
\end{subfigure}
\begin{subfigure}{0.325\textwidth}
\centering
\includegraphics[width=1\linewidth]{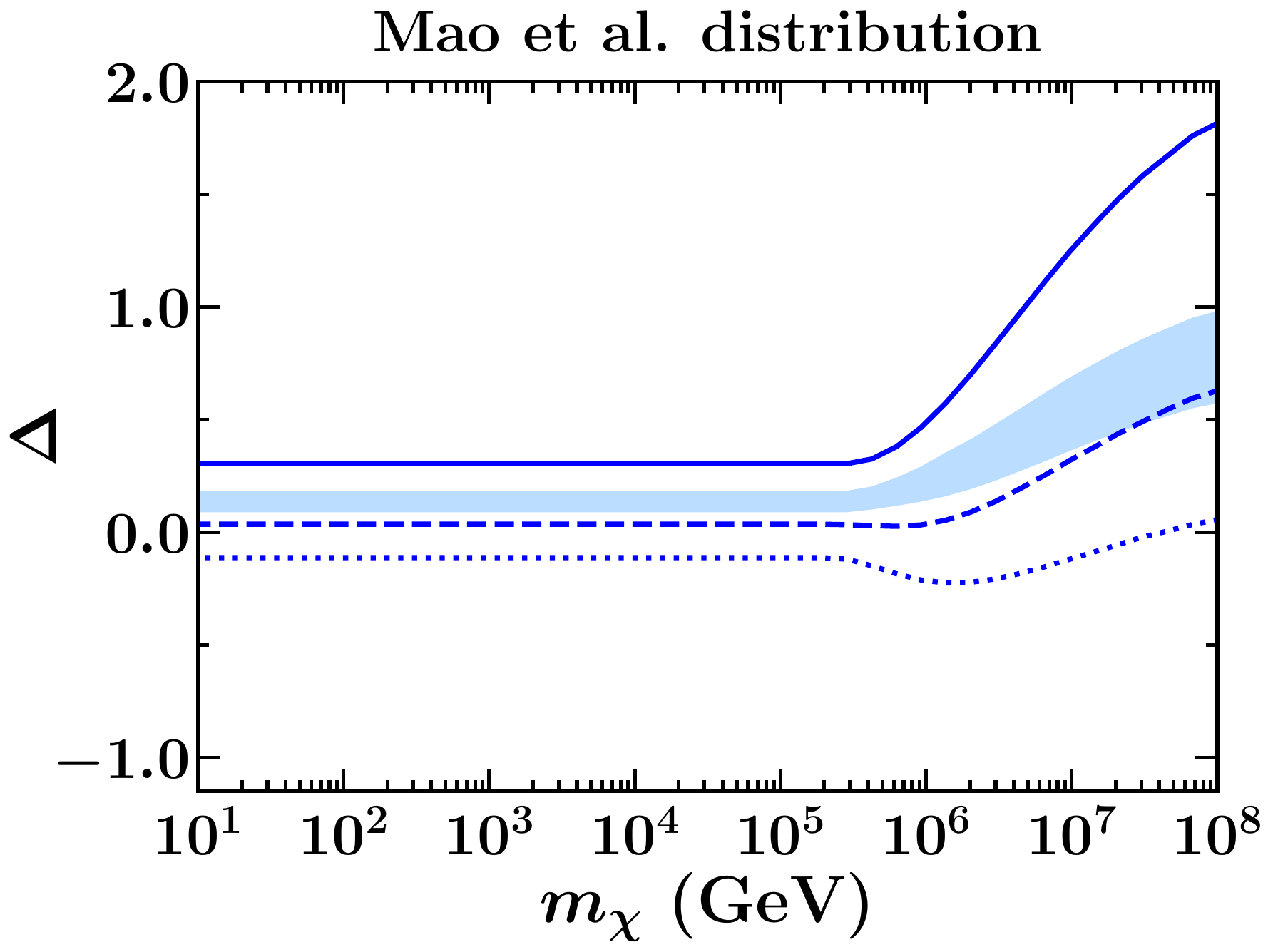} 
\caption{}
\label{sf:Mao1}
\end{subfigure}
\caption{The relative change in the DM capture rate using Eq. \eqref{eq:capt-effective}, for a neutron star of mass $1.5 \, {\rm M_{\odot}}$ and radius $10 \, {\rm km}$. The DM-nucleon scattering cross-section is fixed at $10^{-45} \, {\rm cm^2}$. The individual plots show the relative change in capture rate for MB, King, DPL, Tsallis and the Mao \emph{et al.} distribution, given by magenta, yellow, red, green and blue colors respectively. The shaded band in each plot refers to the variation in the relative capture rate by taking into account the astrophysical uncertainty in the estimates of $u_0$ and $u_{\rm esc}$. The solid, dashed and dotted lines show the variations for the ARTEMIS, APOSTLE and FIRE-2 simulations respectively.}
\label{fig:NS}
\end{figure}

In figure \ref{fig:NS}, we present the estimated changes in the capture rate due to the uncertainties and non-MB distributions introduced in section \ref{sec:velocity_dis}, relative to the benchmark values of MB and equipping Eq. \eqref{eq:capt-effective}.
The relative changes for SHM, King distribution, DPL distribution, Mao \emph{et al.} distribution and the Tsallis distribution are shown by the magenta, yellow, red, green and blue colored plots respectively. The solid, dashed and dotted curves denote the relative changes introduced by fitting the velocity distribution parameters to the ARTEMIS, APOSTLE and FIRE-2 haloes respectively. The shaded bands show the variations in the relative capture rate for $u_0=233 \pm 3$ and $u_{\rm esc}=528_{-25}^{+24}$, evaluated from astrophysical  observations. For the distributions considered here, we note that the largest departures in capture rates, with respect to the capture rate evaluated at the benchmark values of MB is due to the Mao \emph{et al.} velocity distribution. In the range of astrophysical uncertainty, we report a maximum change of $98 \%$  in the capture rate for the Mao \emph{et al.} distribution. Similarly, if we consider $u_0$ and $u_{\rm esc}$ from cosmological simulations, we find the APOSTLE and FIRE-2 curves to lie in close proximity of the SHM\footnote{The value of $u_0$ for the APOSTLE lie close to the central value of the astrophysically measured quantities. Hence, the capture rate computed using APOSTLE data show least departure from the capture rate obtained from the benchmark values of SHM. From a similar argument, capture rate estimated from the ARTEMIS simulation shows maximum deviation from SHM. The effects due to a change in $u_0$ although similar, exceeds the effects due to changes in $u_{\rm esc}$.}. We have evaluated the parameters relevant to this work, for the FIRE-2 simulation, which is additional to ARTEMIS and APOSTLE considered in \cite{Maity:2020wic}. In addition, among the three simulations considered here, FIRE-2 has slightly finer mass and spacial resolutions and considers accretions at the solar circle while incorporating mesh-free hydrodynamics. In case of the NS considered here, we report the corresponding value to be $22\%$ from the estimated parameters of FIRE-2 simulation provided in Table \ref{tab:bestfit}. The relative changes estimated for each distribution for the FIRE-2 simulation is given in Table \ref{tab:PerUnc}. On a similar note, the deviations obtained for the APOSTLE and ARTEMIS simulations are at $63\%$ and $181\%$ respectively.

As can be seen from Eq. \eqref{eq:caprate}, the velocity distribution leaves a direct footprint on the capture rates. In Fig. \ref{sf:comp} we plot the velocity distributions for the different velocity profiles considered, using the benchmark values of $u_0$ and $u_{\rm esc}$. Unlike SHM where a sharp cut-off is introduced in the velocity distribution at the galactic escape velocity, the non-SHM distributions gradually attain zero at $u_{\rm esc}$. Therefore, the non-SHM distributions do not over-predict the number of particles present in their tails. Hence one would expect a reduced capture rate for the non-SHM distributions like the King, DPL, Mao \emph{et al.} and Tsallis relative to SHM.

Lower the initial velocity, more likely it is for the particles to get captured. Therefore, it is the low velocity tail of the velocity distributions which are sensitive to the DM capture rate. An increase in the DM velocity dispersion would result in a flattened profile. Consequently, more number of DM particles can be found in the low and high velocity tails of the distribution, accompanied by a reduction in the number of particles near the most probable velocity. This getting reflected through Eq. \eqref{eq:caprate}, would result in a reduction of the DM captured rate. The effect although similar, is quiet less due to an increment in the galactic escape velocity. This explains the variations in capture rate, when computed using the best fit values from different simulations for a particular distribution.

In Fig. \ref{fig:NS}, we find the relative change in the capture rate to be insignificant and almost constant at low  masses of DM. It is only above $\sim 2 \times 10^5$ GeV that the relative change with respect to the benchmark SHM increases appreciably. For the massive particles which have velocities close to the most probable velocity, would require multiple scatters before they can be captured by the NS. For the typical NS that has been considered here, a massive DM particle of mass $100$ PeV moving with an initial velocity $ \sim 90 \,\rm km/s$, requires $\sim 10$ scatters before its velocity becomes less than the escape velocity of the NS. This is why the multi-scatter framework becomes important in order to probe higher  orders of DM mass. This effect has been qualitatively shown in figure \ref{sf:scat} and discussed in appendix \ref{sec:app3}. In Fig. \ref{sf:comp} we plot the velocity distributions upto $\sim 90 \,\rm km/s$, upto which we get a significant contribution of the distributions to the capture rate. For the observationally motivated values, we find that the area under the curve near the small velocity tail of the Mao \emph{et al.} distribution being maximum, provides the strongest deviation from the SHM capture rate. Followed by the Tsallis, DPL and King distributions.

The uncertainties relating to the capture and detection of old NSs have been discussed in \cite{Chatterjee:2022dhp}. For a range of NS equation of states, the capture rate can vary upto $ 40 \%$ having a mass of $1.5 \, \rm M_{\odot}$. However, variations in the NS mass can be accounted to be around $ 10 \%$, which translates to a $ 35 \%$ change in the DM capture rate \cite{Zhu:2019oax}. The most surveyed impact of DM capture inside old NSs is the study of their heating signatures. Nevertheless, NSs observed in the UV regime are not ideal for probing dark heating signatures \cite{Kargaltsev:2003eb,deLavallaz:2010wp}. For an infrared telescope like the James Webb Space Telescope (JWST),  its Near-Infrared Camera (NIRCAM) can potentially detect such heating signatures owing to the capture of DM  \cite{Baryakhtar:2017dbj,Chatterjee:2022dhp}. The maximum uncertainty in the NS capture rate translate to a $66\%$ change in the projected limits of DM-nucleon scattering cross-section, considering the potential detection of a NS with surface temperature of $1600$K \cite{Joglekar:2020liw,Maity:2021fxw}.

\subsection{White dwarfs}
\label{sec:wd}

White dwarfs (WD) are the stellar-core remnant of a low mass star, which are not massive enough to initiate later stages of burning at their cores. Such a star balances its gravitational collapse by its electron degeneracy pressure. A typical white dwarf is mostly comprised of oxygen and carbon, however the dominant contribution comes from carbon nuclei. 

We consider a typical carbon dominated WD of unit solar mass extending upto a radius of $10^4 \, {\rm Km}$. We consider a DM-nucleon scattering cross-section of the order of $10^{-38} \, \rm cm^2$.

\begin{figure}[t]
\centering
\begin{subfigure}{0.325\textwidth}
\centering
\includegraphics[width=1\linewidth]{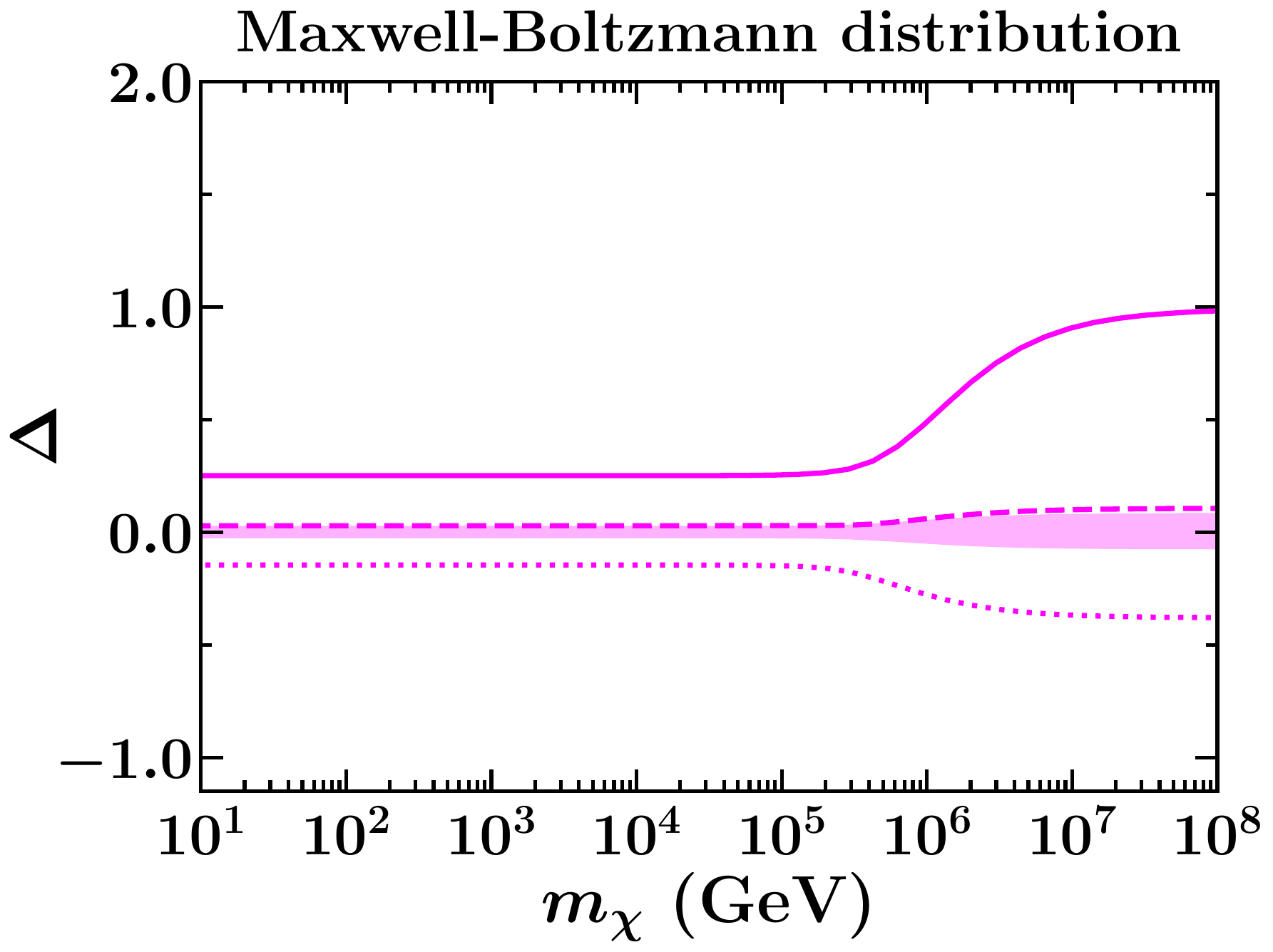} 
\caption{}
\label{sf:MB2}
\end{subfigure}
\begin{subfigure}{0.325\textwidth}
\centering
\includegraphics[width=1\linewidth]{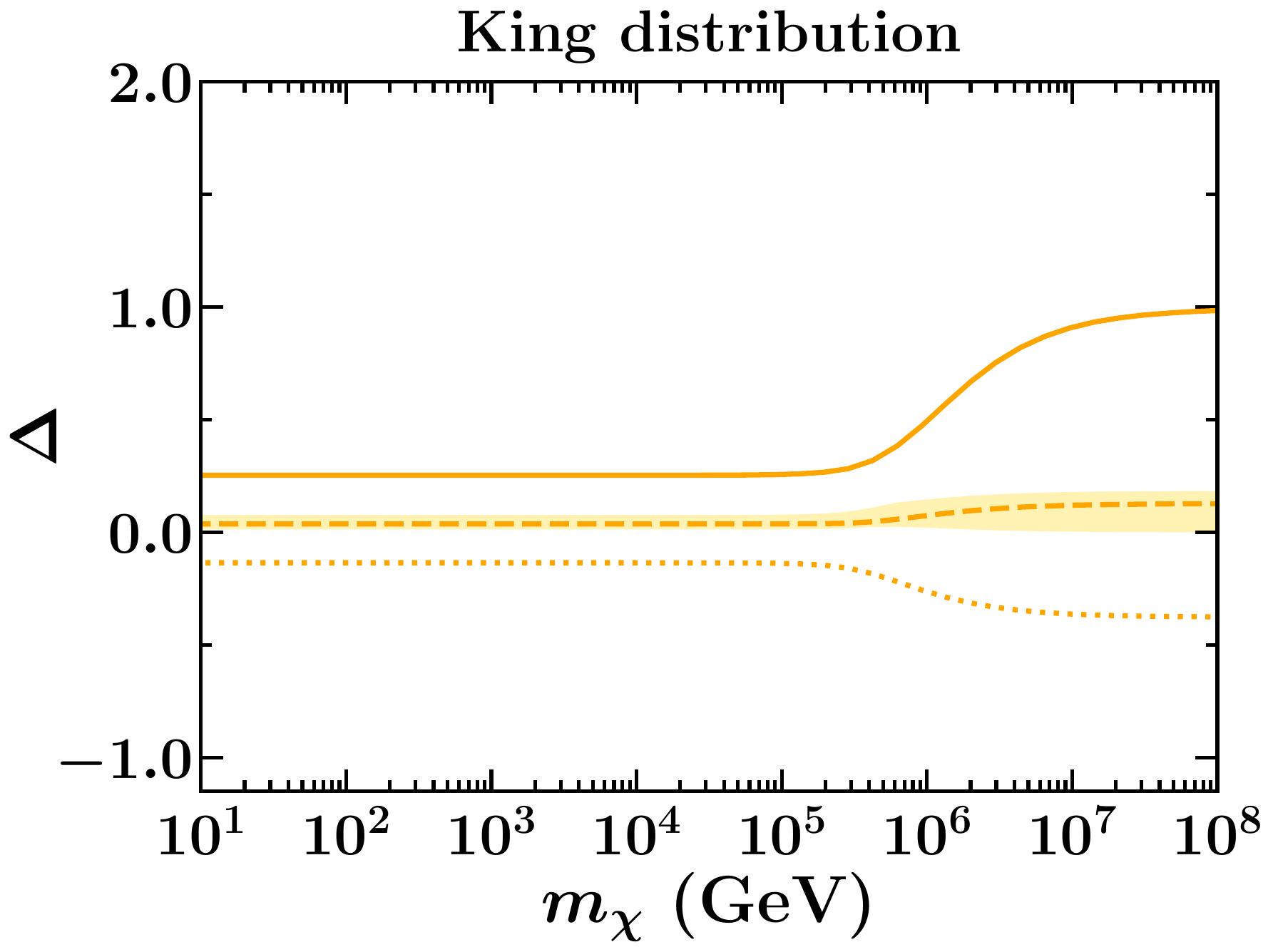} 
\caption{}
\label{sf:King2}
\end{subfigure}
\begin{subfigure}{0.325\textwidth}
\centering
\includegraphics[width=1\linewidth]{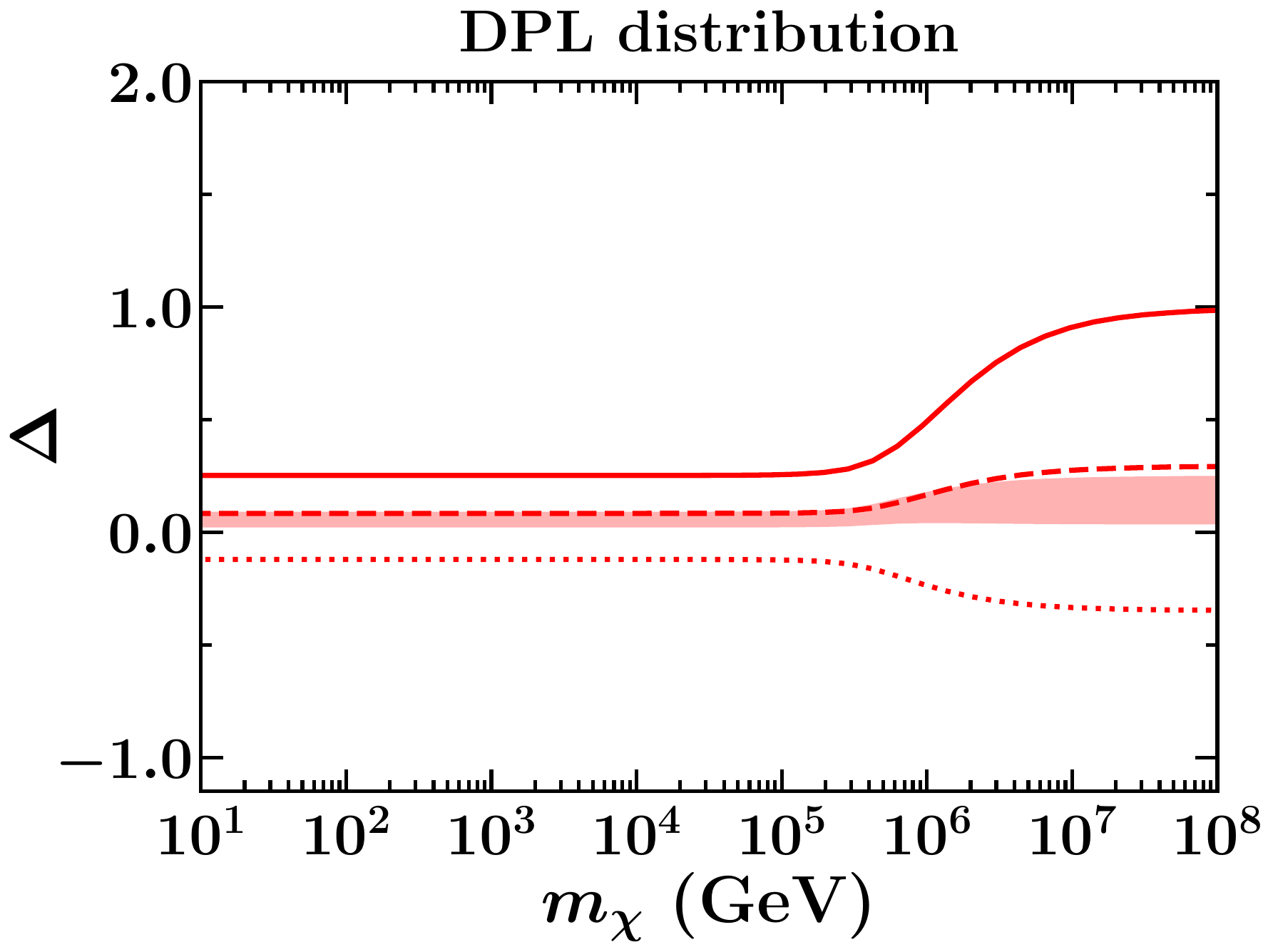} 
\caption{}
\label{sf:DPL2}
\end{subfigure}
\begin{subfigure}{0.325\textwidth}
\centering
\includegraphics[width=1\linewidth]{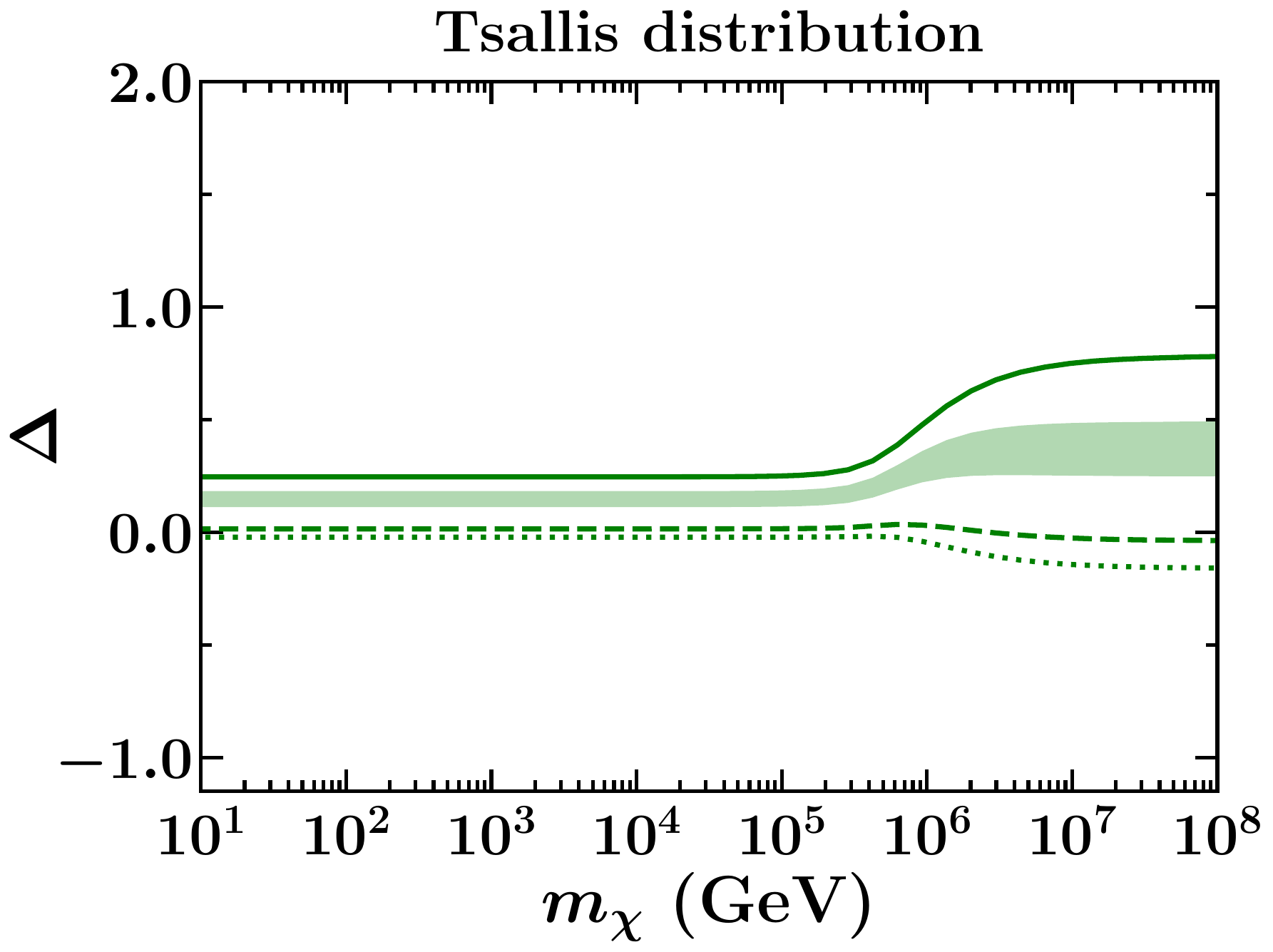} 
\caption{}
\label{sf:Tsal2}
\end{subfigure}
\begin{subfigure}{0.325\textwidth}
\centering
\hspace{0.35em} \vspace{0.55em}
\includegraphics[width=0.65\linewidth]{figs/index} 
\caption*{}
\end{subfigure}
\begin{subfigure}{0.325\textwidth}
\centering
\includegraphics[width=1\linewidth]{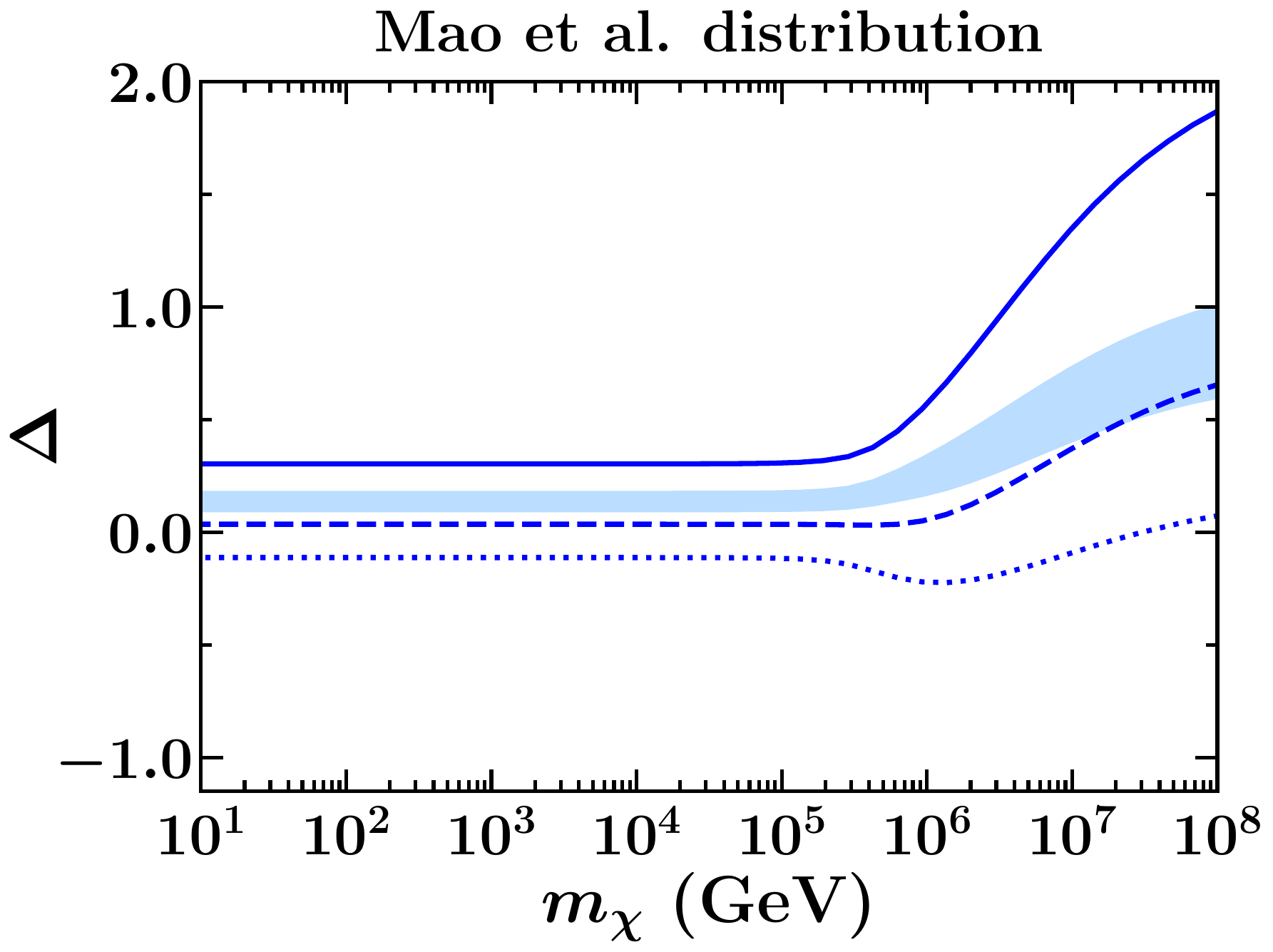} 
\caption{}
\label{sf:Mao2}
\end{subfigure}
\caption{Same as Fig. \ref{fig:NS}, but for a white dwarf of mass $1 \, {\rm M_{\odot}}$ and radius $10^4 \, {\rm km}$. The DM-nucleon scattering cross-section is fixed at $10^{-38} \, {\rm cm^2}$.}
\label{fig:WD}
\end{figure}

In Fig. \ref{fig:WD} we plot the relative changes in the capture rate calculated from Eq. \eqref{eq:capt-effective} for the different velocity distributions, similar to Fig. \ref{fig:NS}. We report a change of $100\%$ in the capture rate taking into the uncertainties introduced from observations, for the Mao \emph{et al.} velocity distribution. Whereas, we find a corresponding $22\%$ change in the capture rate using the best fit parameters for the FIRE-2 simulation, provided in Table \ref{tab:bestfit}. The variations in the DM capture rate for the individual distributions taken up for this work are summarized in Table \ref{tab:PerUnc}. For APOSTLE and ARTEMIS we find a corresponding deviation of $ 65\% $ and $186\% $ respectively. Similar to our study of NS, we notice the multi-scatter framework to set in at a DM mass of $\sim 10^5$ GeV, from where the change in the relative capture rate increases significantly.

Apart from the uncertainties discussed in this work, the ambiguities while determining the mass and radius of a WD would lead to similar but reduced effects in the capture rate. The errors obtained in determining the mass and radius of a unit solar mass WD is found to be $1\%$ and $3\%$ respectively \cite{Bond2017TheSS,Holberg_1998}. These errors can lead to a maximum $17\%$ variation in the DM capture rate. Assuming the high DM density at the M4 globular cluster, projected bounds on the DM-nucleon scattering cross-section have been derived from the heating signatures of old WDs \cite{McCullough:2010ai,Dasgupta:2019juq,Bell:2021fye}. The uncertainty in determining the luminosity quoted in \cite{Bedin2009THEEO} would translate to an uncertainty of only $2\%$ in the WD effective temperature. Assuming the DM dispersion velocity to be $8 \, {\rm km/s}$ and DM density to be $798 \, {\rm GeV \, cm^{-3}}$ at the M4 cluster \cite{McCullough:2010ai,Bell:2021fye}, the projected limits on DM-nucleon scattering cross-section can vary upto $124\%$ for Mao \emph{et al.} distribution, for DM mass of $10^3 \, {\rm GeV}$.
\subsection{Brown dwarfs}
\label{sec:bd}

Brown dwarfs are sub-stellar objects which lie in between the mass range of planets and stars, with masses about $15-75$ times that of Jupiter and a radius comparable to that of Jupiter. The density of BDs although greater than that of planets, are not dense enough to start Hydrogen burning in their cores. This makes them ideal to search for heating signatures resulting from the  capture of DM.

\begin{figure}[t]
\centering
\begin{subfigure}{0.325\textwidth}
\centering
\includegraphics[width=1\linewidth]{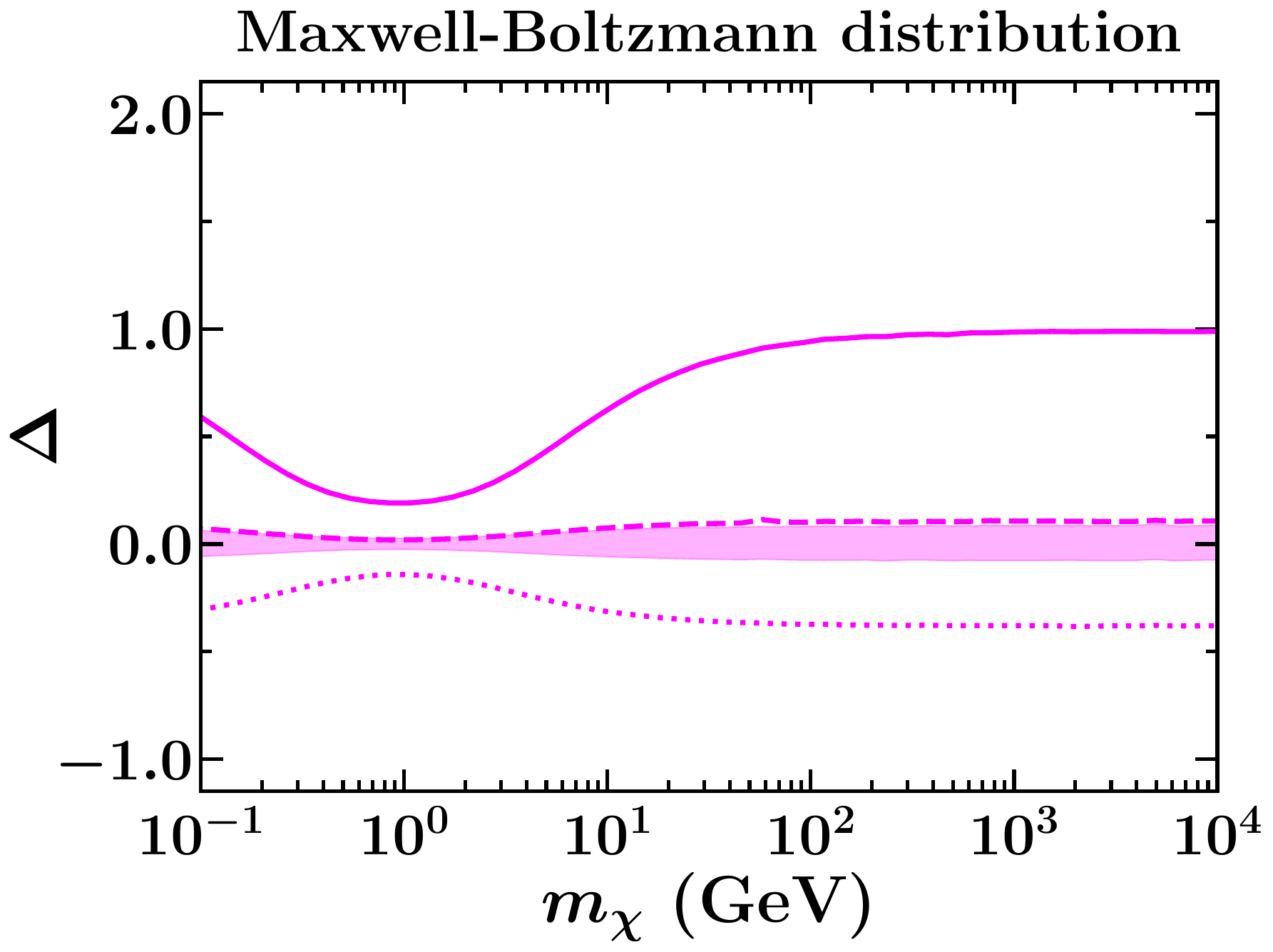} 
\caption{}
\label{sf:MB3}
\end{subfigure}
\begin{subfigure}{0.325\textwidth}
\centering
\includegraphics[width=1\linewidth]{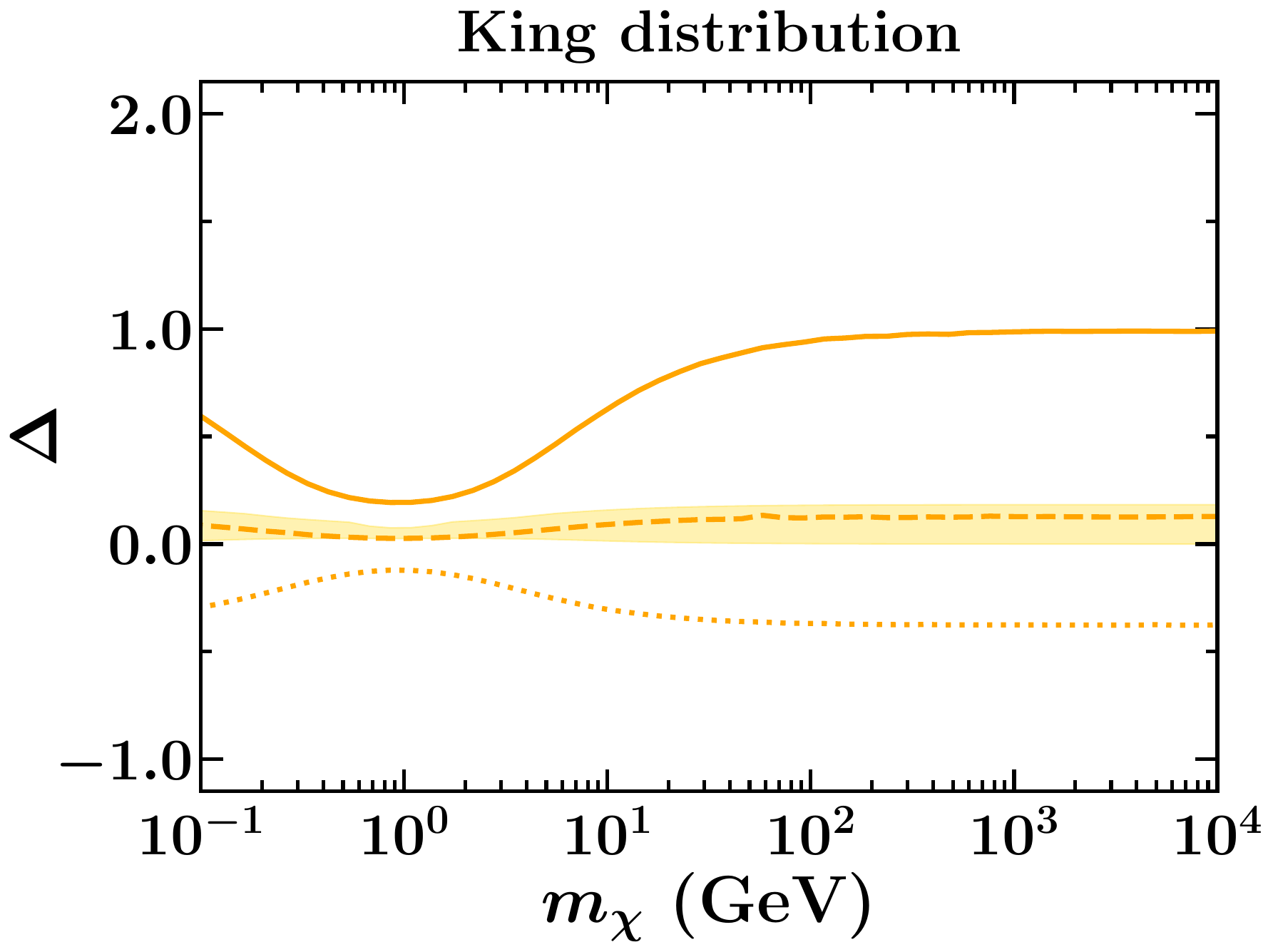} 
\caption{}
\label{sf:King3}
\end{subfigure}
\begin{subfigure}{0.325\textwidth}
\centering
\includegraphics[width=1\linewidth]{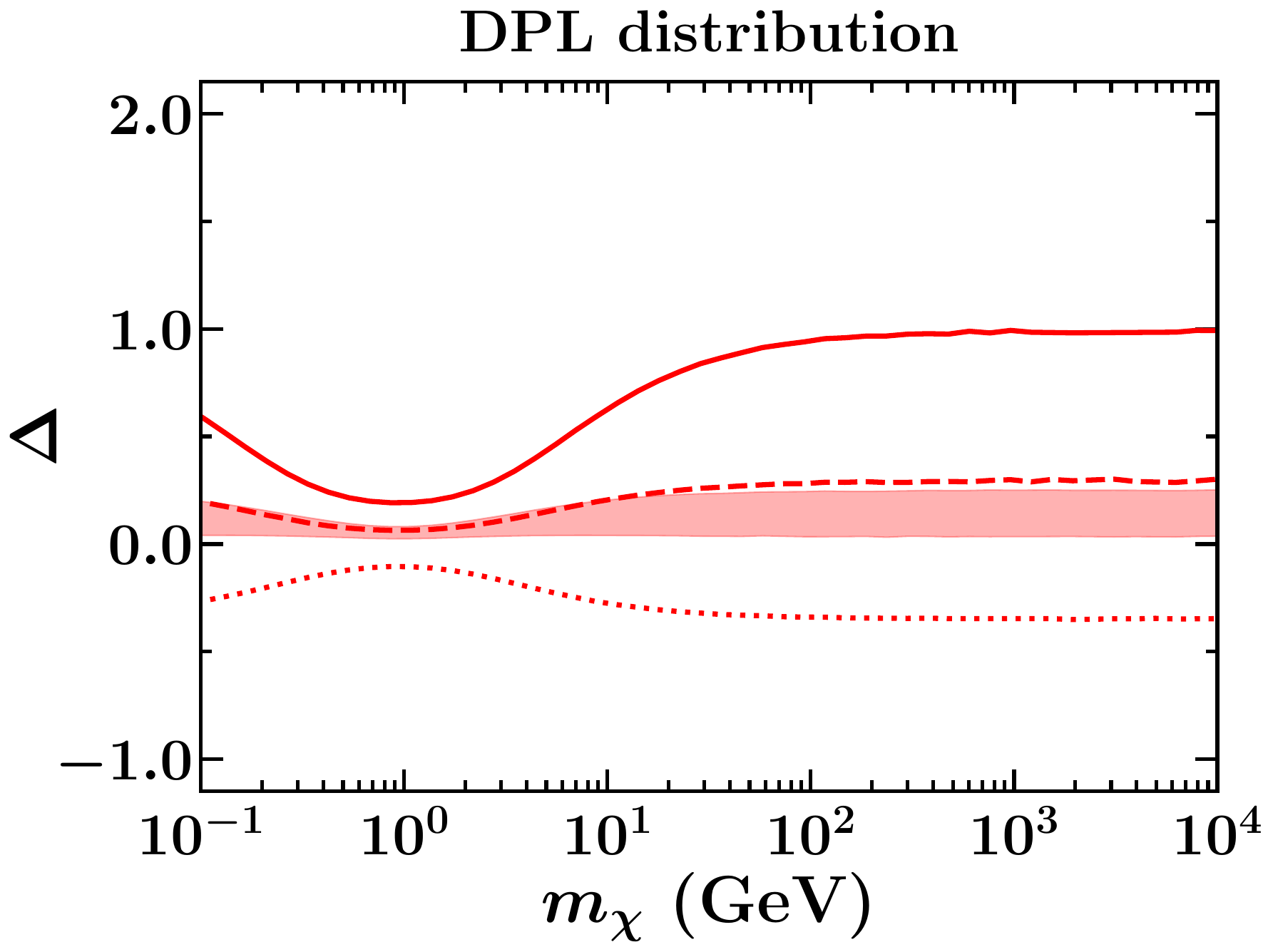} 
\caption{}
\label{sf:DPL3}
\end{subfigure}
\begin{subfigure}{0.325\textwidth}
\centering
\includegraphics[width=1\linewidth]{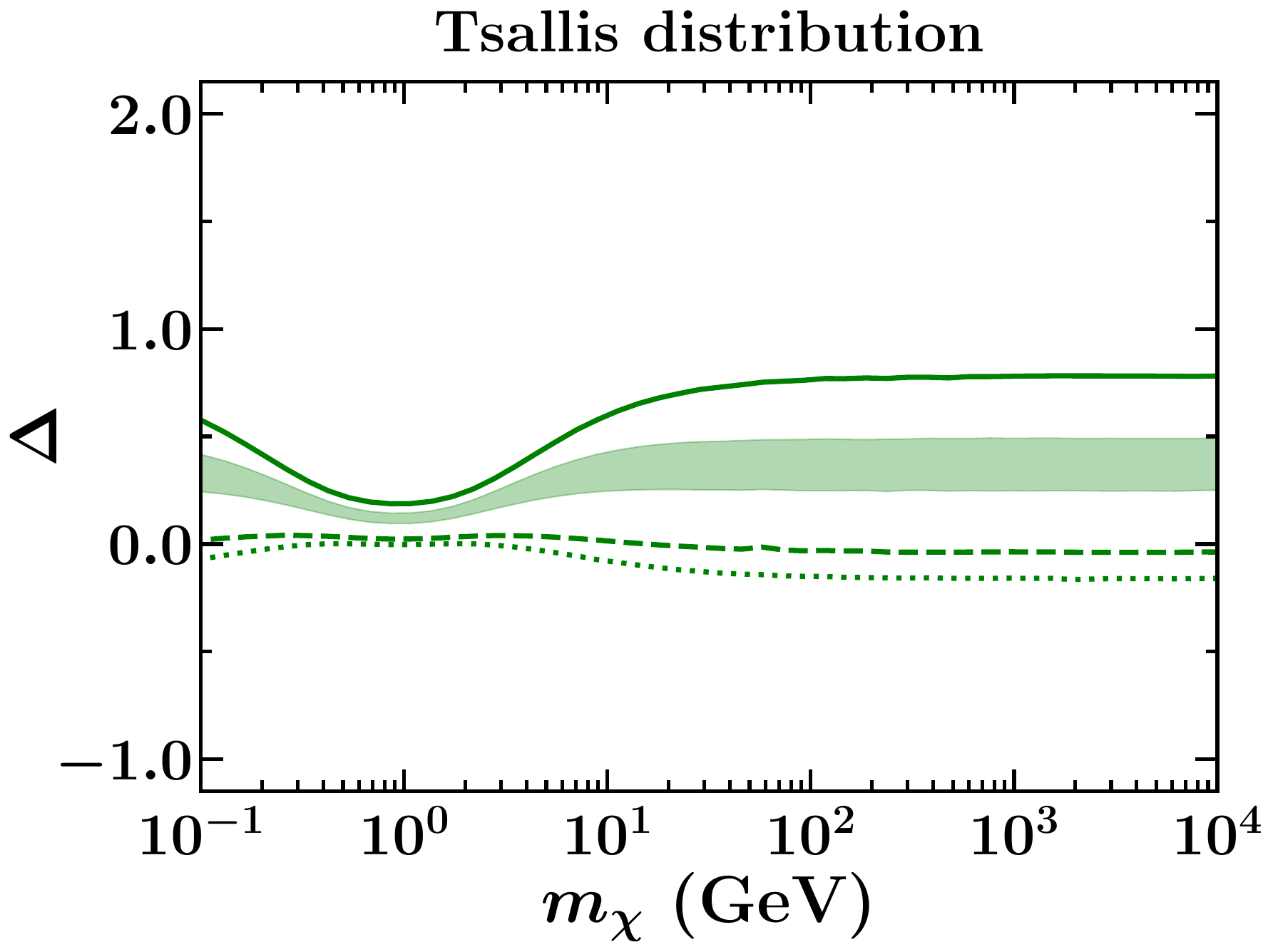} 
\caption{}
\label{sf:Tsal3}
\end{subfigure}
\begin{subfigure}{0.325\textwidth}
\centering
\hspace{0.35em} \vspace{0.55em}
\includegraphics[width=0.65\linewidth]{figs/index} 
\caption*{}
\end{subfigure}
\begin{subfigure}{0.325\textwidth}
\centering
\includegraphics[width=1\linewidth]{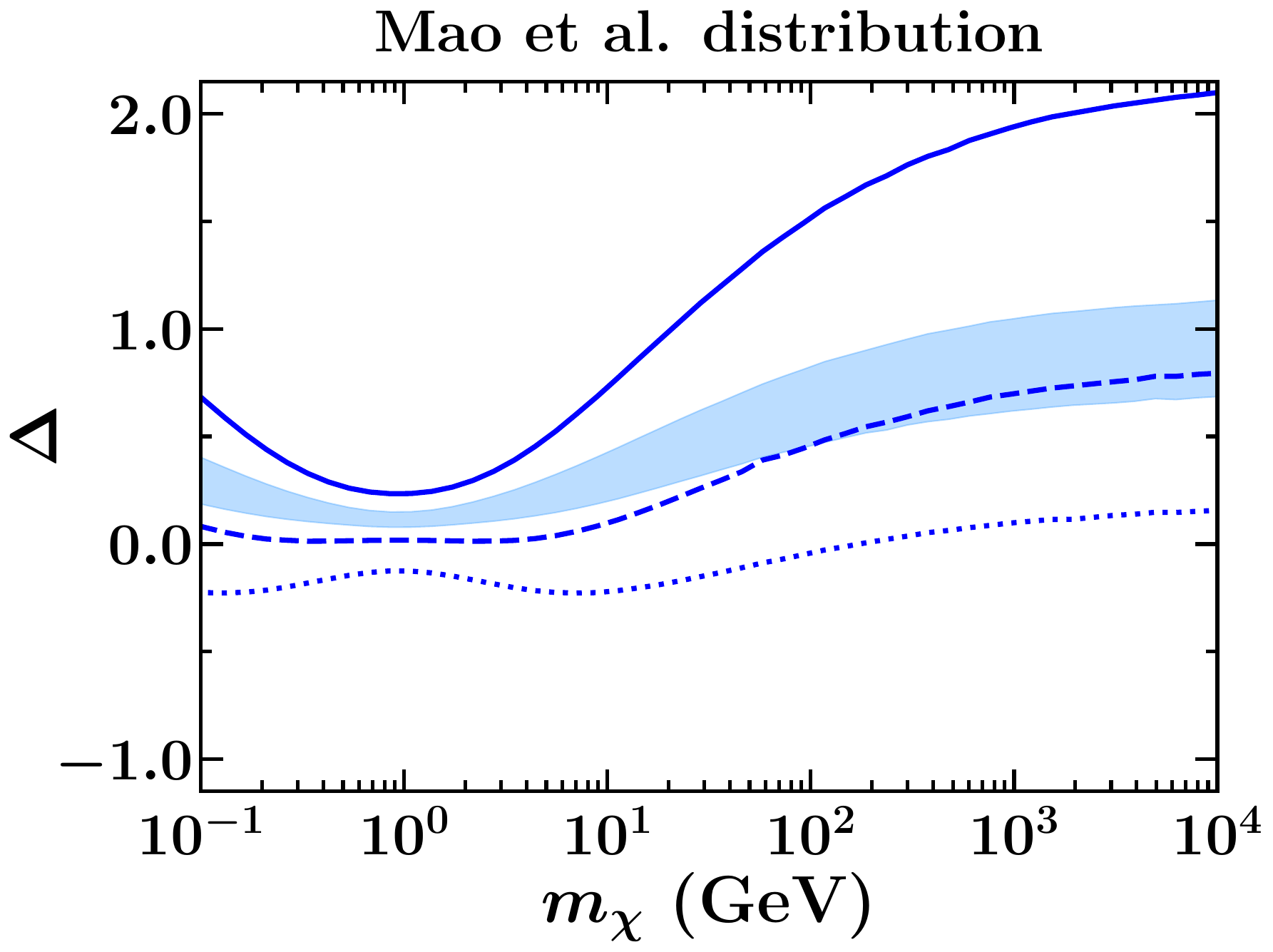} 
\caption{}
\label{sf:Mao3}
\end{subfigure}
\caption{Same as Fig. \ref{fig:NS}, but for a brown dwarf of mass and radius $55 \, {\rm M_{\rm J}}$ and ${\rm R_{\rm J}}$ respectively. We have considered a DM-nucleon scattering cross-section at $10^{-36} \, {\rm cm^2}$.} 
\label{fig:BD}
\end{figure}


We consider a DM-nucleon scattering cross-section of $ 10^{-36} \, \rm cm^2 $ in BDs of the size of Jupiter (${\rm R_{\rm J}}$), having a mass $55$ times that of Jupiter (${\rm M_{\rm J}}$).

Taking into account the observational uncertainties, we report a maximum change of $113\%$ in the DM capture rate for the Mao \emph{et al.} velocity distribution, using Eq. \eqref{eq:capt-effective}. A similar change in the capture rate can be identified at $22\%$  by considering the best fit values from FIRE-2 simulation. The values of which are tabulated in Table \ref{tab:bestfit}. For the APOSTLE and ARTEMIS simulations, we report a maximum variation in the capture rate at $79\%$ and $209\%$ respectively. Figure \ref{fig:BD} shows the variation in the capture rate for the different velocity distributions, similar to Fig. \ref{fig:NS}.

To understand the effect of the limitations related to the mass and radius of BDs, we utilize the values quoted in \cite{Beatty2018}. The observational errors are $8\%$ and $7\%$ respectively for the mass and radius of a BD. This renders a maximum $40\%$ variation in the capture rate, which is less than the variations in capture rate predicted in this work. The exoplanets with effective temperatures less than $500$ K can be measured through the Mid-Infrared Instrument (MIRI), whereas the ones with effective temperature above $500$ K can be measured with the Near-Infrared Imager and Slitless Spectrograph (NIRISS) of the JWST \cite{Leane:2020wob}. Considering the possible detection of $\mathcal{O}(100)$K BD \cite{Leane:2020wob}, the projected limit on DM-nucleon scattering cross-section can vary upto $67\%$ due to the maximum uncertainty in capture rate.

\subsection{Exoplanets}
\label{sec:exo}

Planets are the smallest known celestial objects heavy enough for their self-gravity to give them a spherical shape. Those orbiting a star and having cleared the neighborhood around their orbits are termed as planets. Though the term exoplanet is used for the planets which are not a part of our solar system, the analysis we have followed can also be applied to the gaseous planets present within our solar system. The milky way galaxy is estimated to host a billion stars. Jupiter like gaseous planets have a high density and a large radius, which makes them ideal candidates for the study of DM capture \cite{geosciences8100362,Leane:2020wob}. In comparison to NSs, the temperatures of exoplanets can be measured upto large distances from the galactic center, for which they can probe DM-density dependent heating signals with higher significance and lower exposure time. 

\begin{figure}[t]
\centering
\begin{subfigure}{0.325\textwidth}
\centering
\includegraphics[width=1\linewidth]{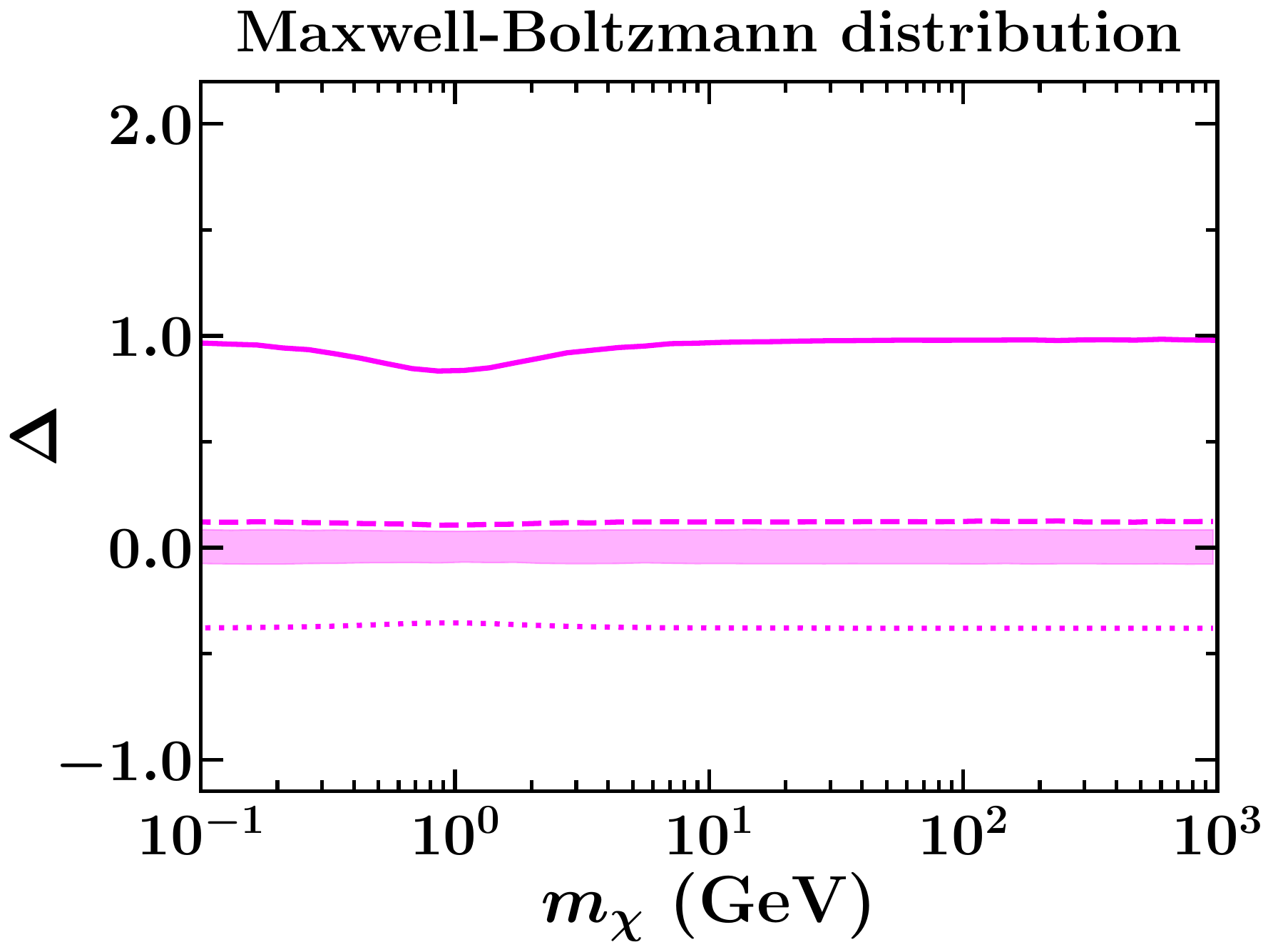} 
\caption{}
\label{sf:MB4}
\end{subfigure}
\begin{subfigure}{0.325\textwidth}
\centering
\includegraphics[width=1\linewidth]{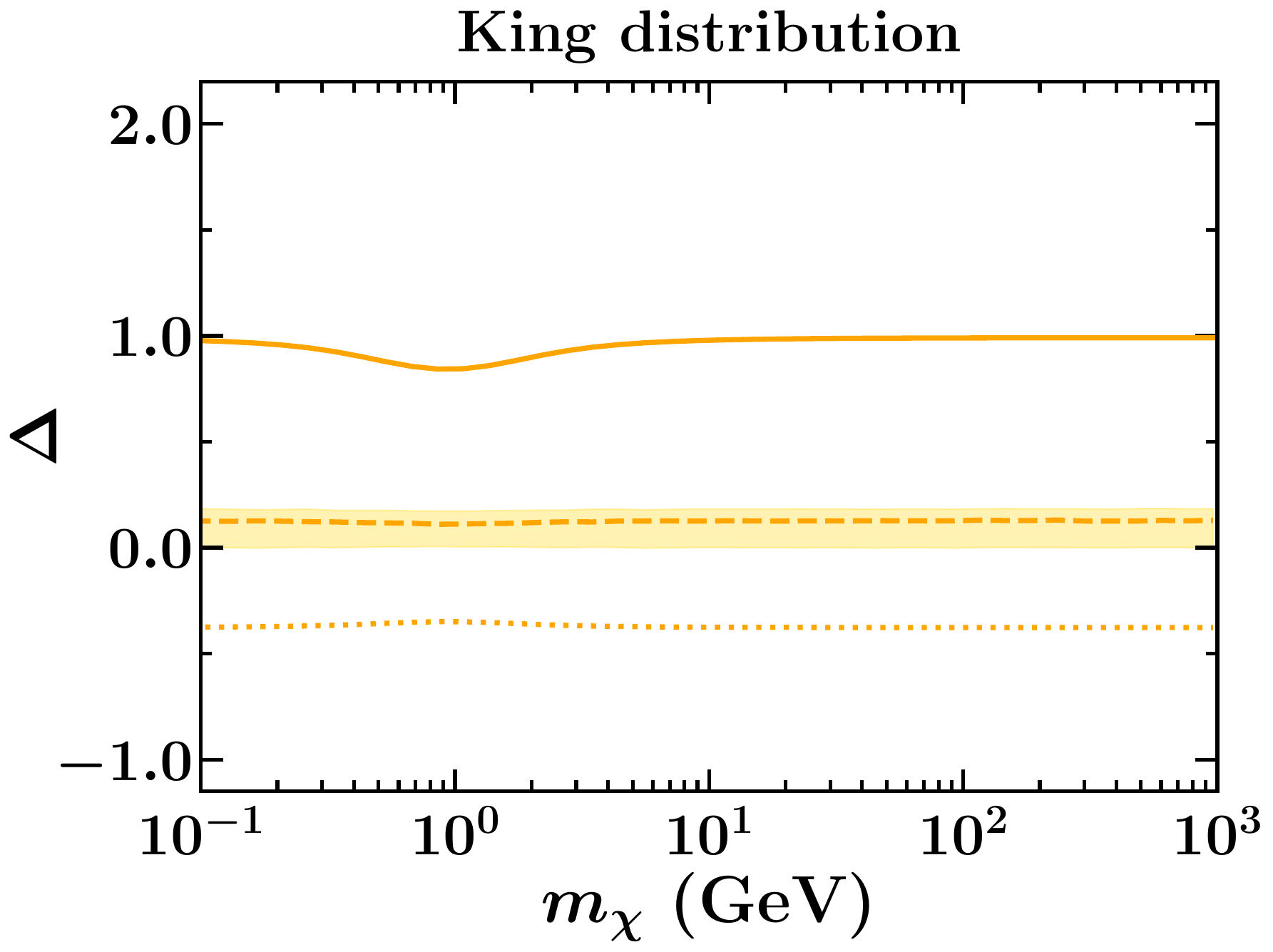} 
\caption{}
\label{sf:King4}
\end{subfigure}
\begin{subfigure}{0.325\textwidth}
\centering
\includegraphics[width=1\linewidth]{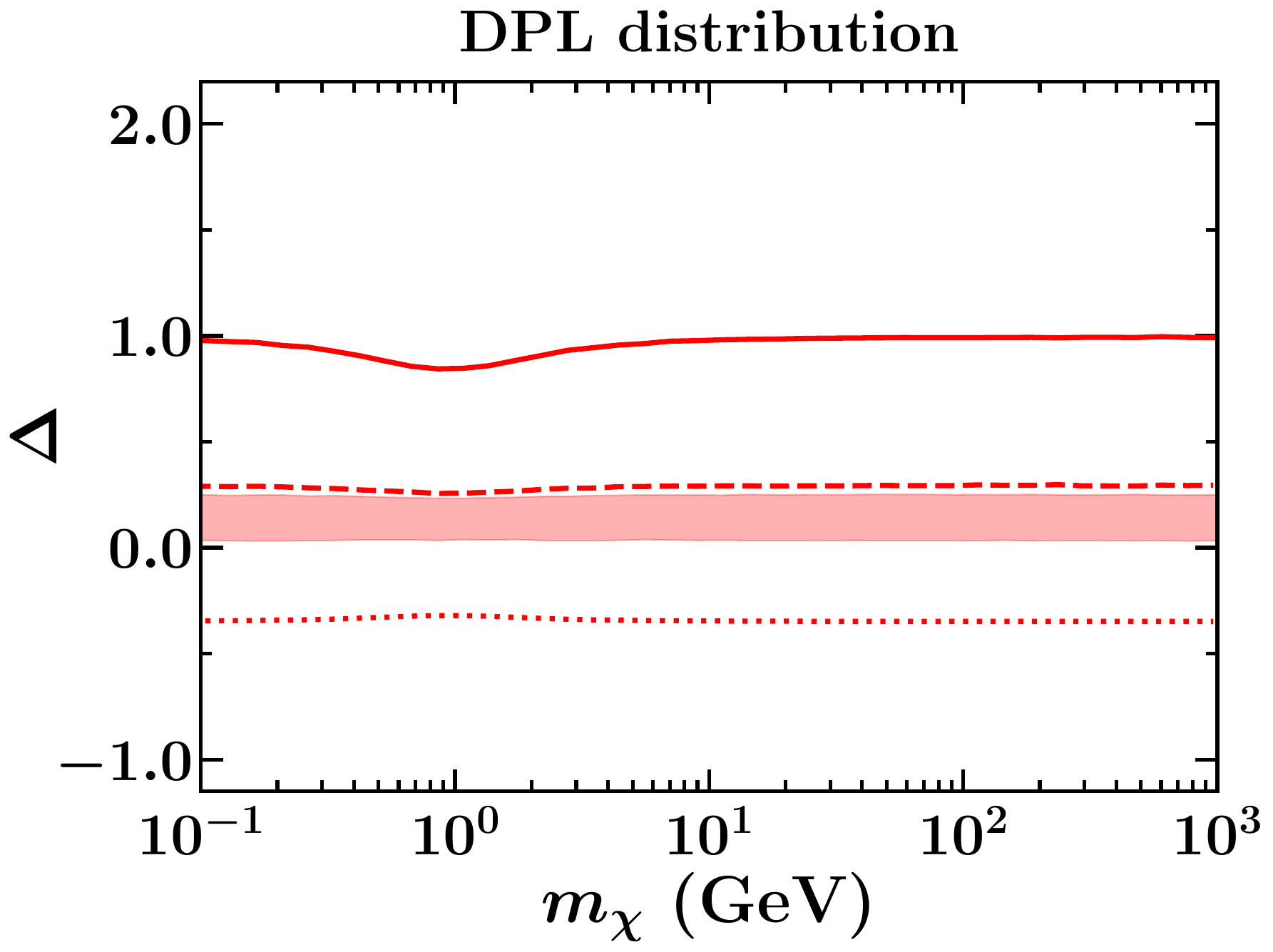} 
\caption{}
\label{sf:DPL4}
\end{subfigure}
\begin{subfigure}{0.325\textwidth}
\centering
\includegraphics[width=1\linewidth]{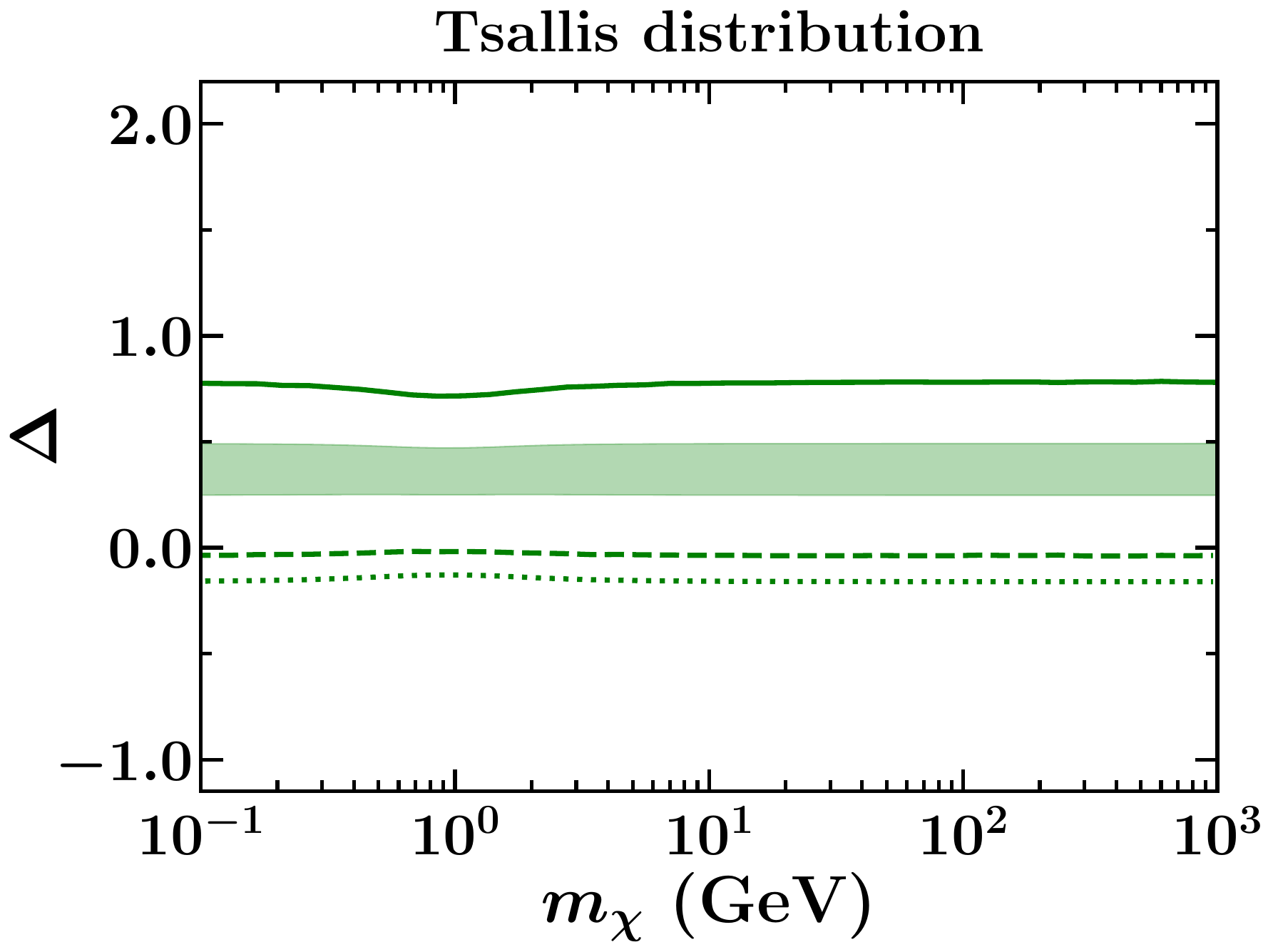} 
\caption{}
\label{sf:Tsal4}
\end{subfigure}
\begin{subfigure}{0.325\textwidth}
\centering
\hspace{0.35em} \vspace{0.55em}
\includegraphics[width=0.65\linewidth]{figs/index} 
\caption*{}
\end{subfigure}
\begin{subfigure}{0.325\textwidth}
\centering
\includegraphics[width=1\linewidth]{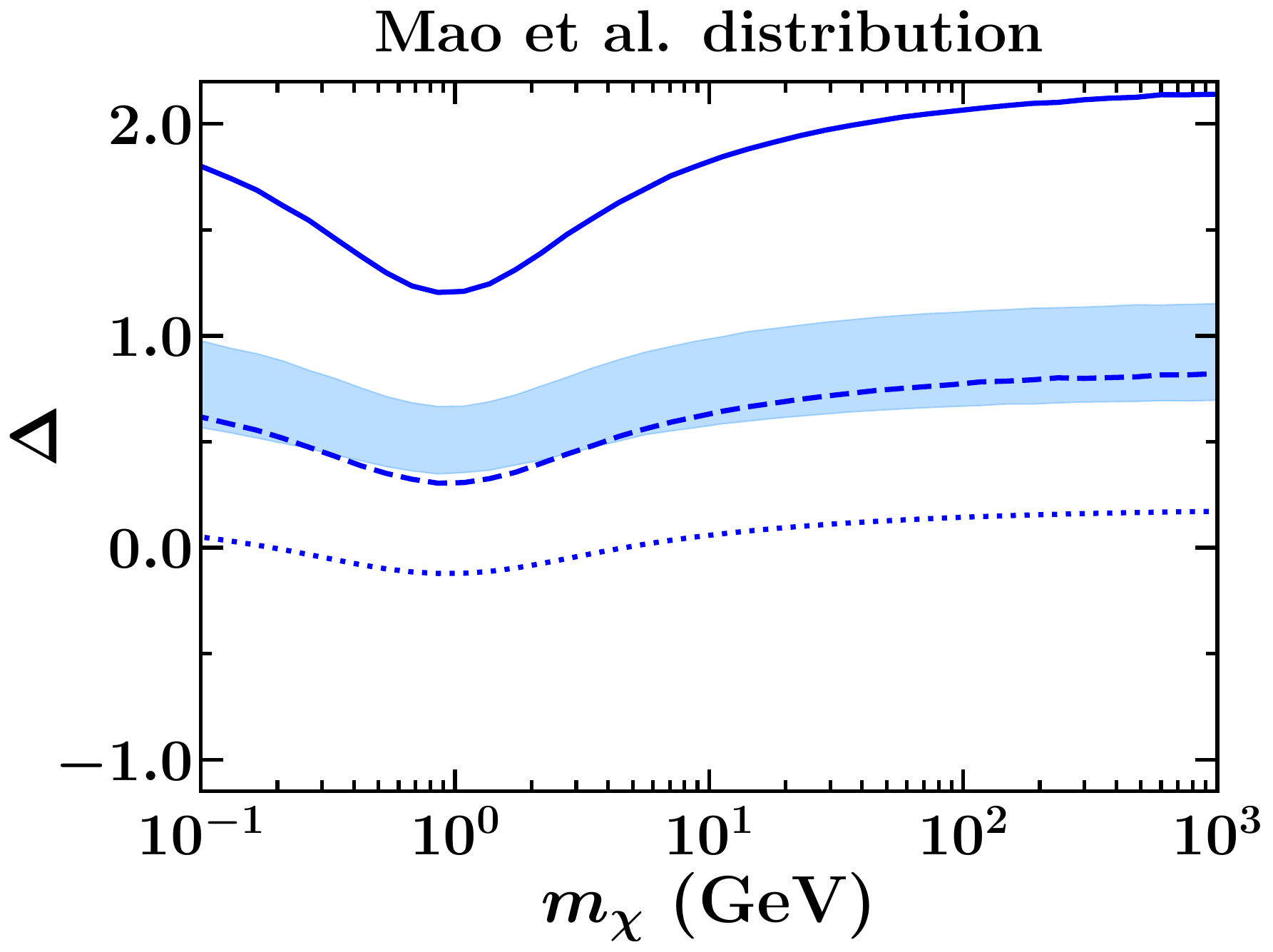} 
\caption{}
\label{sf:Mao4}
\end{subfigure}
\caption{Same as Fig. \ref{fig:NS}, but for an exoplanet of mass ${\rm M_{\rm J}}$ and radius ${\rm R_{\rm J}}$. The DM-nucleon scattering cross-section is fixed at $10^{-34} \, {\rm cm^2}$.}
\label{fig:exo}
\end{figure}


We consider a typical exoplanet of mass and size equal to that of Jupiter and consider a DM-nucleon scattering cross-section of $ 10^{-34} \, \rm cm^2$. As discussed in \cite{Bramante:2022pmn}, the effective escape velocity of a planet also depends on the escape velocity of its host star, evaluated at the planet's position. However, a considerable portion of exoplanets are estimated to have been ejected from their host stars’ planetary environment \cite{Mroz2017,Leane:2020wob}. To keep our analysis generic, we have not assumed any influence from the host star.

For the considered exoplanet we find a maximum deviation of $116\%$ in the capture rate, with respect to the benchmark values of SHM, for the Mao \emph{et al.} velocity distribution for the observational uncertainties. Whereas, if we consider the input parameters from the best fit to FIRE-2 simulation, we find a corresponding deviation of $17\%$. The relative change in the capture rate for the individual distributions are tabulated in Table \ref{tab:PerUnc}. For the APOSTLE and ARTEMIS simulations we obtain a maximum change of $82\%$ and $213\%$ in the capture rate  respectively. The uncertainties arising from different velocity distribution have been shown in Fig. \ref{fig:exo}.

The uncertainties of $3 \%$ and $6 \%$ in determining the radius and mass respectively, of a Jupiter sized exoplanet can translate to a $30\%$ uncertainty in the DM capture rate \cite{kanodia2022toi}. A similar analysis for DM capture within Jupiter has been studied in \cite{Leane:2021tjj}. Jupiter being a part of the solar system has most of its properties precisely measured. Therefore, the uncertainties in measuring the radius and mass of Jupiter are only $\mathcal{O}(10^{-5}\%)$ and they can only produce $\mathcal{O}(10^{-4}\%)$ uncertainty in the capture rate \cite{nasaPlanetaryPhysical}. As discussed in \cite{Leane:2020wob}, the exoplanets with effective temperatures in $\mathcal{O}(100)$ K can be measured through the MIRI and NIRISS of the JWST.  The projected limit on DM-nucleon scattering cross-section, based on a plausible detection of the above mentioned low temperature exoplanet can have an uncertainty of $71\%$ due to the  quoted maximum change in capture rate.

\section{Uncertainty in solar DM capture}
\label{sec:cap_sun}

In this section we will focus on the uncertainties related to DM capture inside our Sun. The general formalism of DM capture discussed in section \ref{sec:formalism} is applicable throughout the class of celestial bodies that can be approximated to have a constant density. Sun being a main sequence star is enriched with elements from hydrogen to iron, hence cannot be approximated to have a constant density. The rate of DM capture within the Sun therefore requires a more rigorous approach. The solar capture rate is given by \cite{Gould:1987ju,Garani:2017jcj,Busoni:2017mhe,Lopes:2020dau},
\begin{equation}
\label{sun_cap}
C_{\odot} = \sum_{i} \left( \frac{\rho_{\chi}}{m_{\chi}} \right) \int_{0}^{R_{\odot}} \, 4 \, \pi \, r^2 \, dr \int_{0}^{u_{\rm esc}} \, du_{\chi} \, \frac{f_{v_{\odot}}(u_{\chi})}{u_{\chi}} \, \left[ u_{\chi}^2 + v_{\rm esc}(r)^2 \right] \int_{0}^{v_{\rm esc}(r)} \, dv \, R_{i}^{-}(w(r) \rightarrow v) \, |F_i(q)|^2.
\end{equation}
Here $v_{\rm esc}$ is the escape velocity, $w(r) = \sqrt{u_{\chi}^2 + v_{\rm esc}(r)^2}$ is the DM velocity at a distance $r$ from the center of the Sun and $R_{\odot}$ denotes the solar radius. $|F_i(q)|^2$ represents the nuclear form factor for the $i$-th target nuclei defined in \cite{Garani:2017jcj,Busoni:2017mhe,Lopes:2020dau}. This form factor turns out to be $1$ if we consider DM scatter with a non-zero spin nucleus like hydrogen. 
In Eq. \eqref{sun_cap}, the term $R_i^{-}(w \rightarrow v)$ defines the rate of scattering of DM with the target nuclei, such that the DM attains a lower velocity $v$ from a velocity $w$ that can be read as \cite{Garani:2017jcj,Lopes:2020dau},
\begin{equation}
\label{R_minus}
R^{-}_i(w(r) \rightarrow v) = \frac{\left( m_{\chi} + m_i \right)^2}{2 \, m_{\chi} \, m_i} \, \frac{n_i(r) \, \sigma_i \, v}{w(r)} \, \Theta\left( v - \left| \frac{m_{\chi} - m_i}{m_{\chi} + m_i} \right| \, w(r) \right).
\end{equation}
Here $n_i(r)$ is the number density of nuclei $i$ at a distance $r$ from the center of the Sun, which has been obtained from \cite{Vinyoles:2016djt}. The velocity distribution profile in Eq. \eqref{sun_cap} is the DM halo velocity distribution measured by an observer at Sun which is given by \cite{Garani:2017jcj},
\begin{equation}
\label{boosted_f}
f_{v_{\odot}}(u_{\chi}) = \frac{1}{2} \, \int_{-1}^{1} \, d \, cos \, \theta_{\odot} \, \, f(\sqrt{u_{\chi}^2 + v_{\odot}^2 + 2 \, u_{\chi} \, v_{\odot} \, cos \, \theta_{\odot}}),
\end{equation}
where $v_{\odot}$ and $\rm cos\theta_{\odot}$ are the solar velocity and the angle between the velocities of the Sun and the DM particle. The rate at which DM gets captured  inside the Sun, depends on the nature of DM and nuclei interaction. For spin dependent (SD) scattering, it is the hydrogen nucleus which primarily contributes to the capture mechanism. However, for the spin independent (SI) scenario, all the elements from helium to iron act as scattering nuclei, which translates to an enhanced capture rate for the SI case.


\begin{figure}[t]
\centering
\begin{subfigure}{0.325\textwidth}
\centering
\includegraphics[width=1\linewidth]{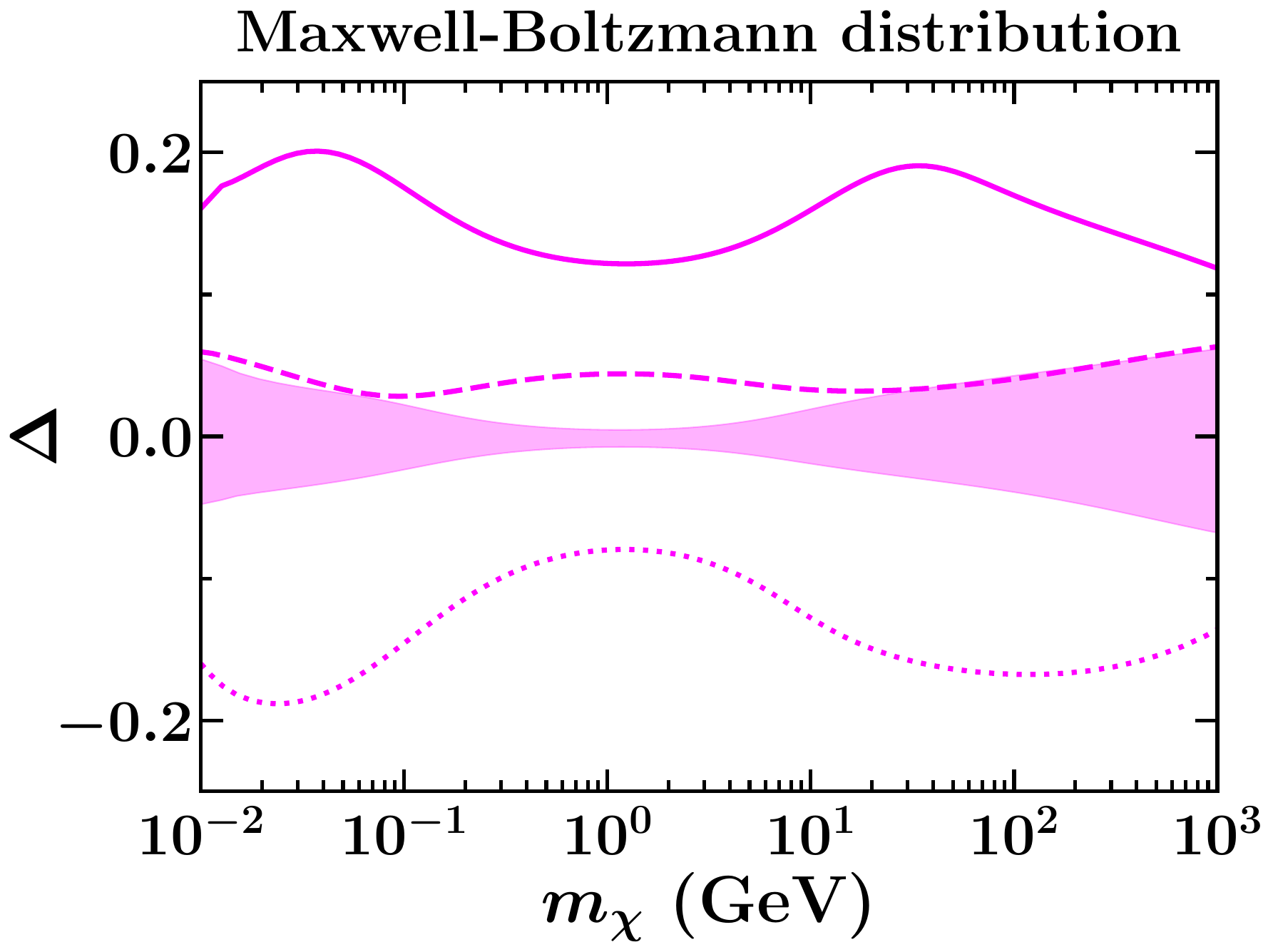} 
\caption{}
\label{sf:MBSD}
\end{subfigure}
\begin{subfigure}{0.325\textwidth}
\centering
\includegraphics[width=1\linewidth]{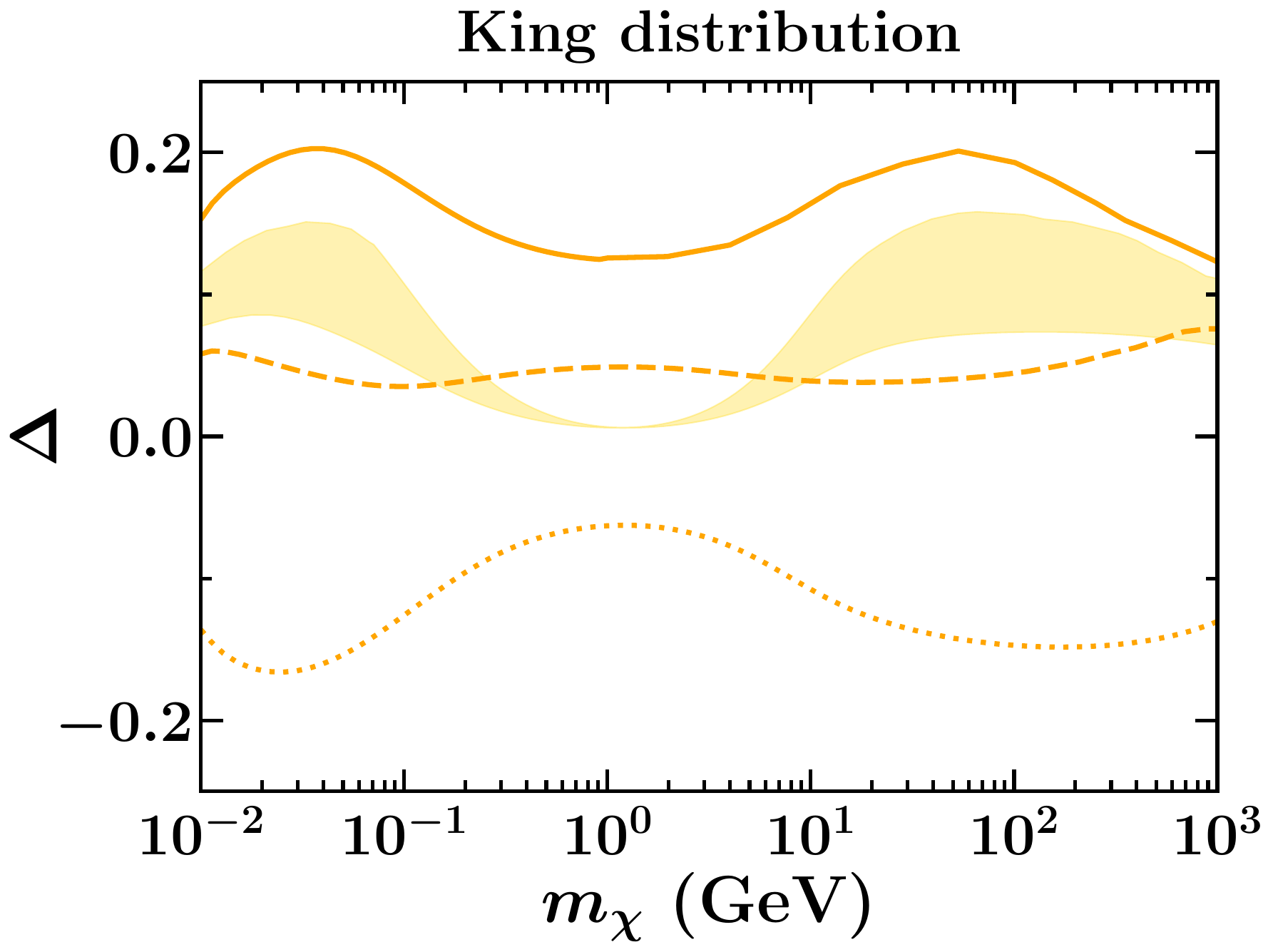} 
\caption{}
\label{sf:KingSD}
\end{subfigure}
\begin{subfigure}{0.325\textwidth}
\centering
\includegraphics[width=1\linewidth]{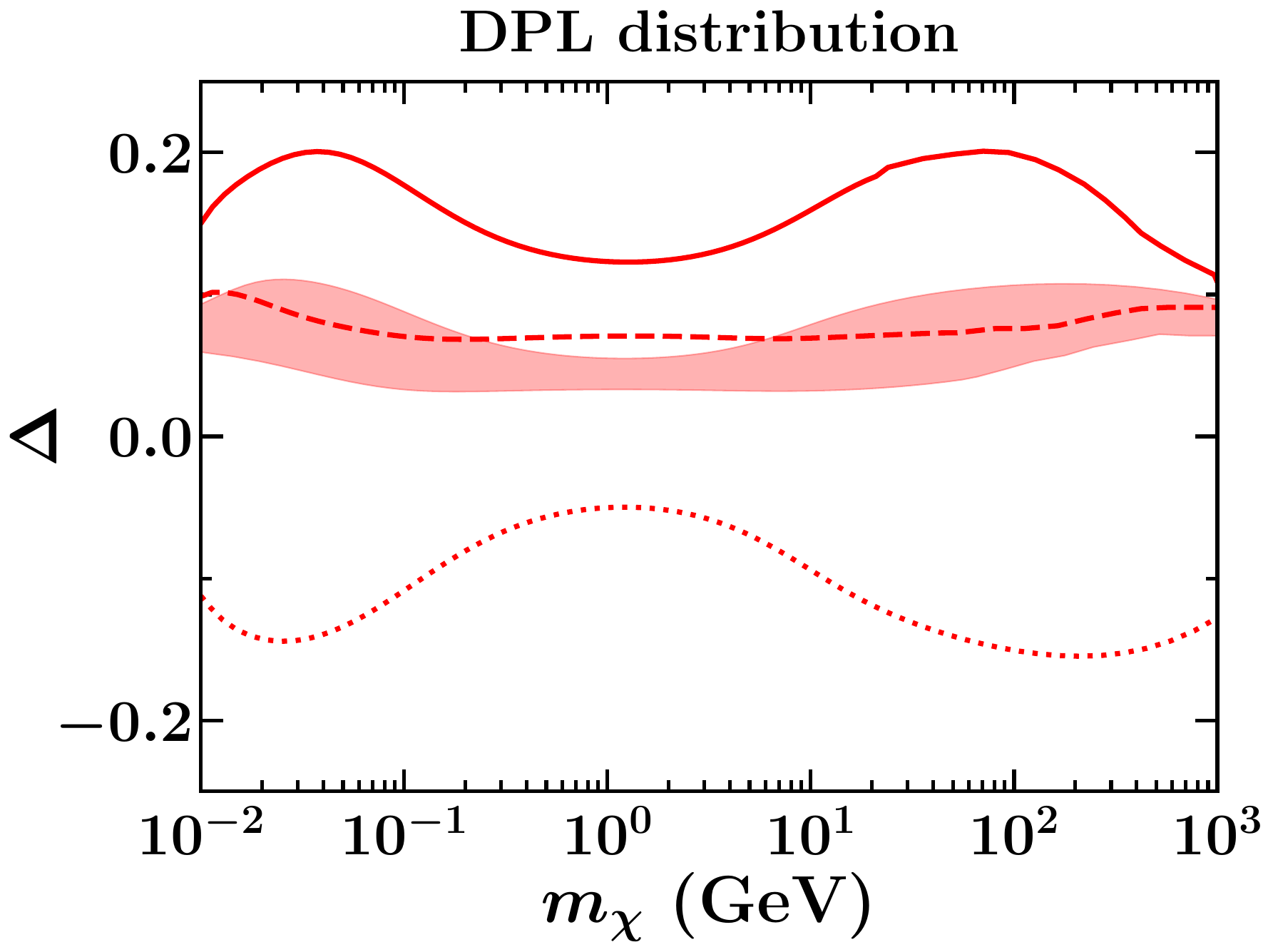} 
\caption{}
\label{sf:DPLSD}
\end{subfigure}
\begin{subfigure}{0.325\textwidth}
\centering
\includegraphics[width=1\linewidth]{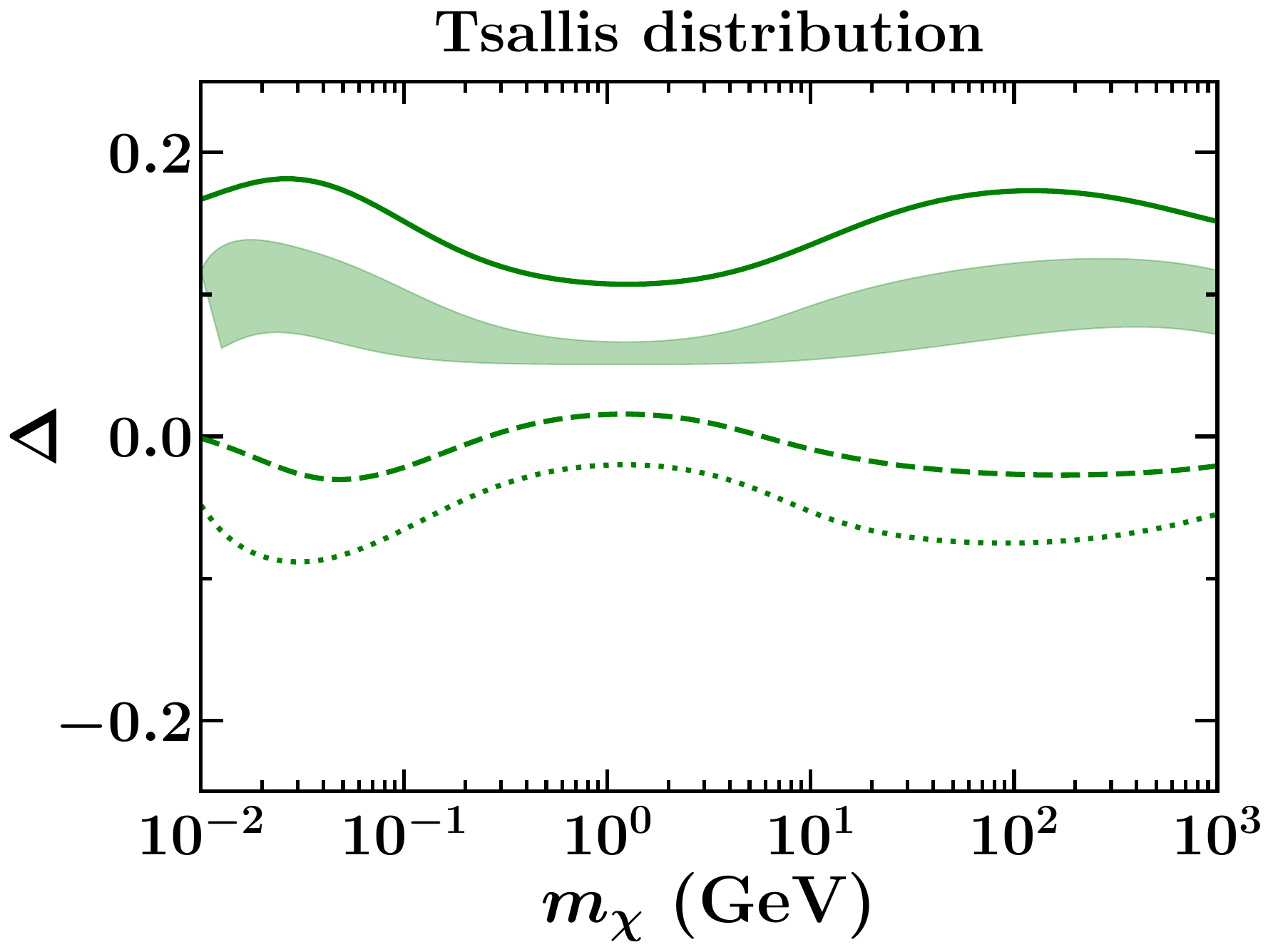} 
\caption{}
\label{sf:TsalSD}
\end{subfigure}
\begin{subfigure}{0.325\textwidth}
\centering
\hspace{0.35em} \vspace{0.55em}
\includegraphics[width=0.65\linewidth]{figs/index}  
\caption*{}
\end{subfigure}
\begin{subfigure}{0.325\textwidth}
\centering
\includegraphics[width=1\linewidth]{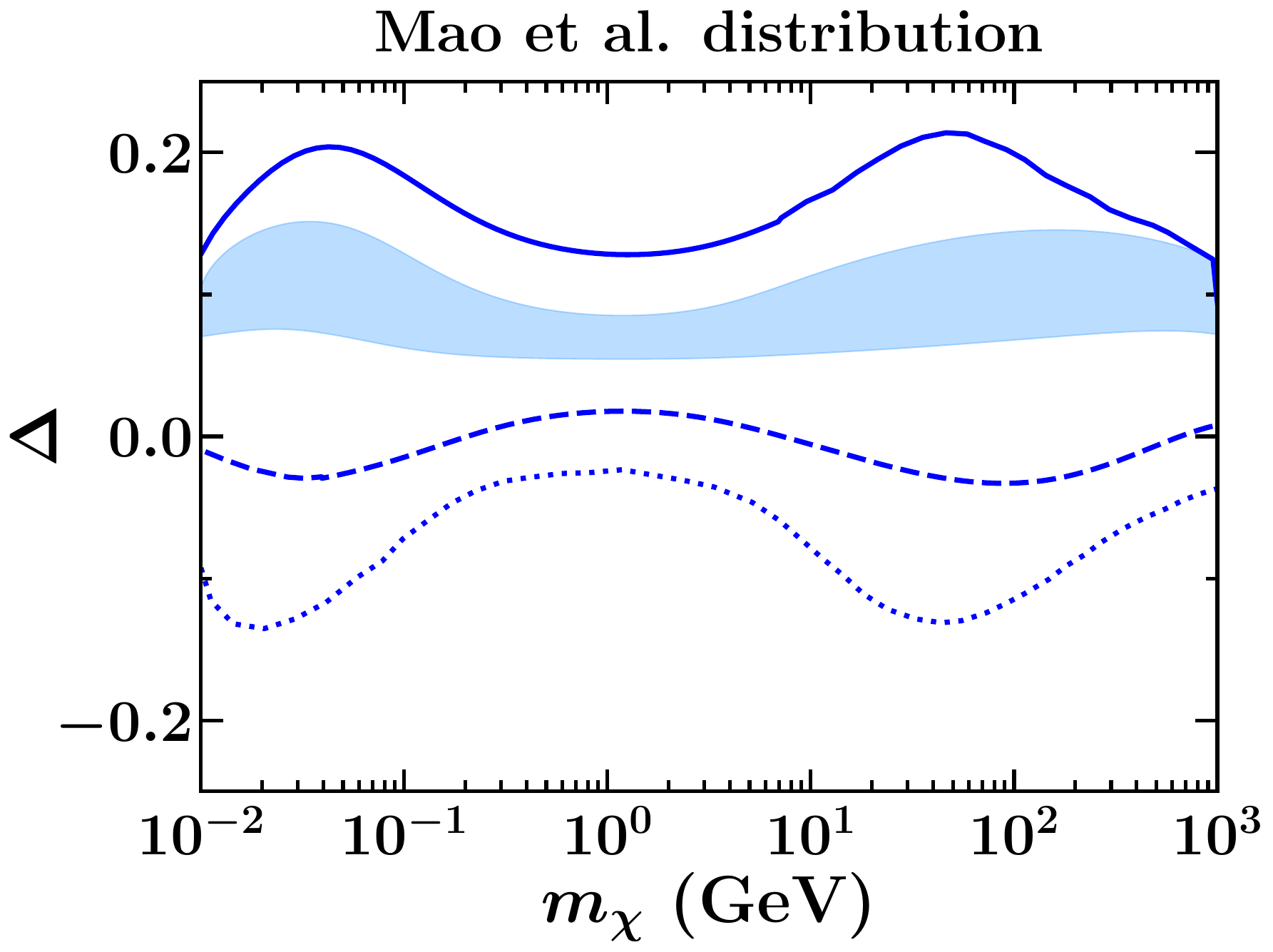} 
\caption{}
\label{sf:MaoSD}
\end{subfigure}
\caption{Same as Fig. \ref{fig:NS}, but for the Spin-Dependent scenario of the Sun. The DM-nucleon scattering cross-section is fixed at $10^{-40} \, {\rm cm^2}$.}
\label{fig:SD}
\end{figure}


\begin{figure}[t]
\centering
\begin{subfigure}{0.325\textwidth}
\centering
\includegraphics[width=1\linewidth]{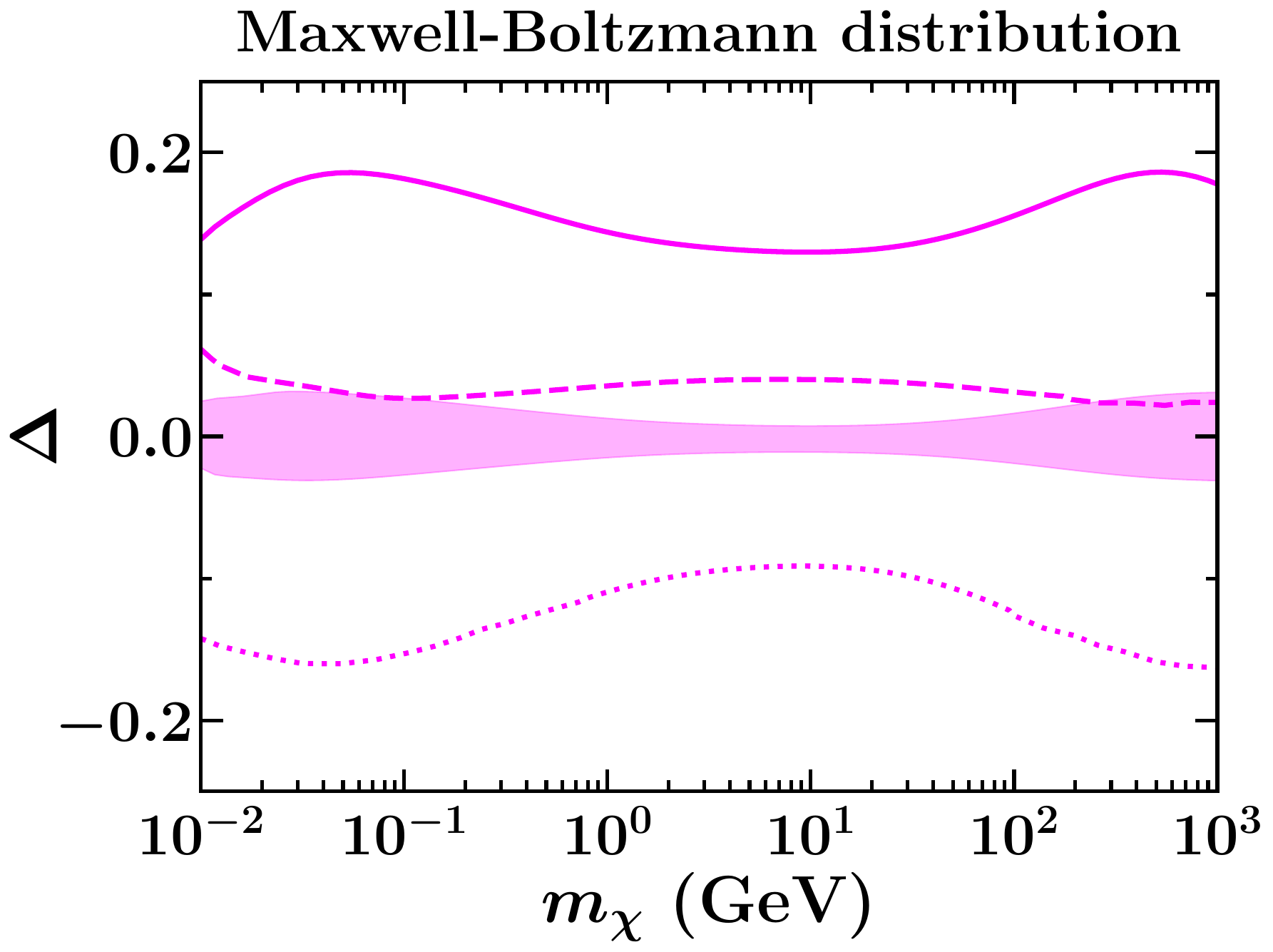} 
\caption{}
\label{sf:MBSI}
\end{subfigure}
\begin{subfigure}{0.325\textwidth}
\centering
\includegraphics[width=1\linewidth]{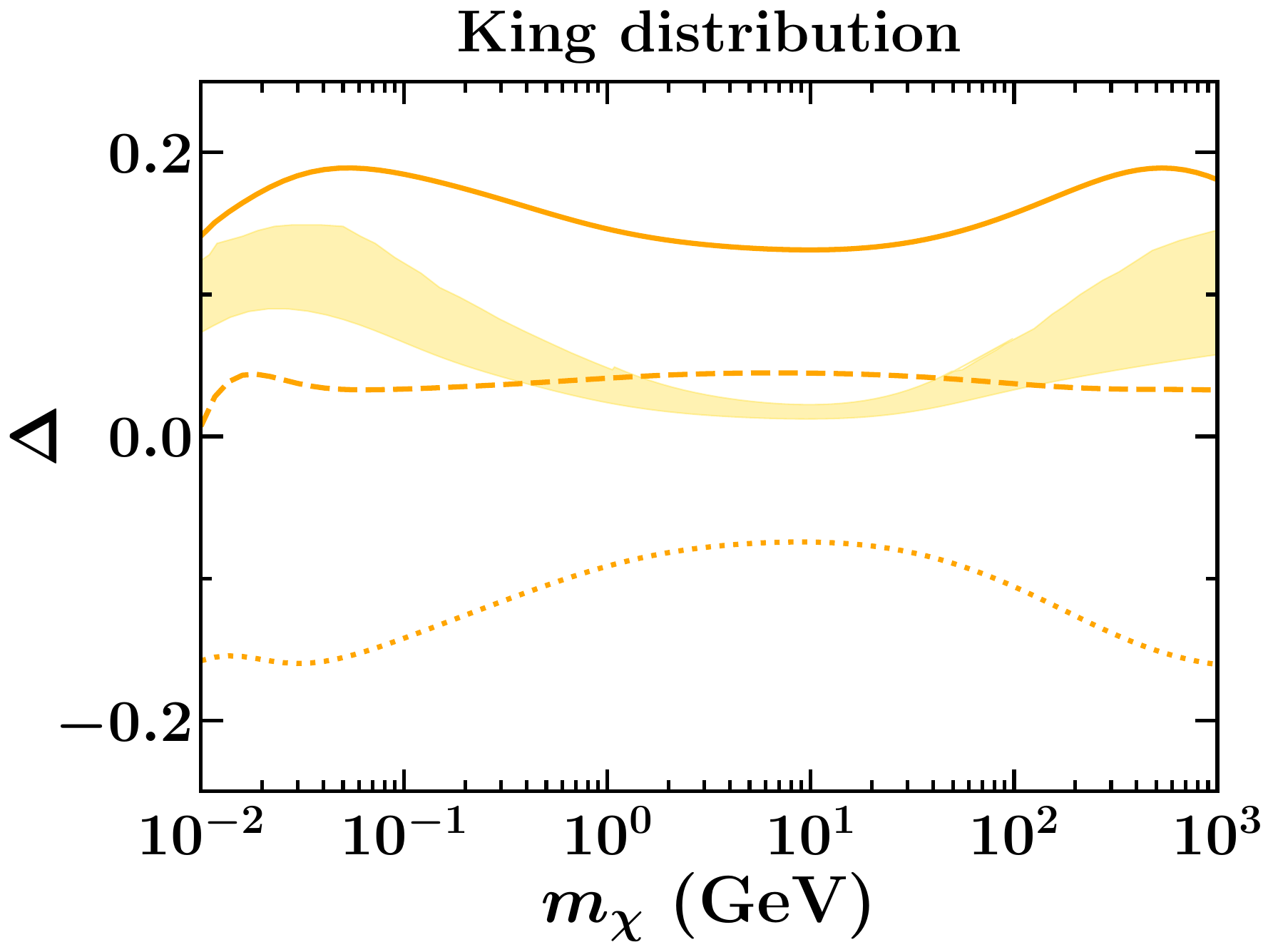} 
\caption{}
\label{sf:KingSI}
\end{subfigure}
\begin{subfigure}{0.325\textwidth}
\centering
\includegraphics[width=1\linewidth]{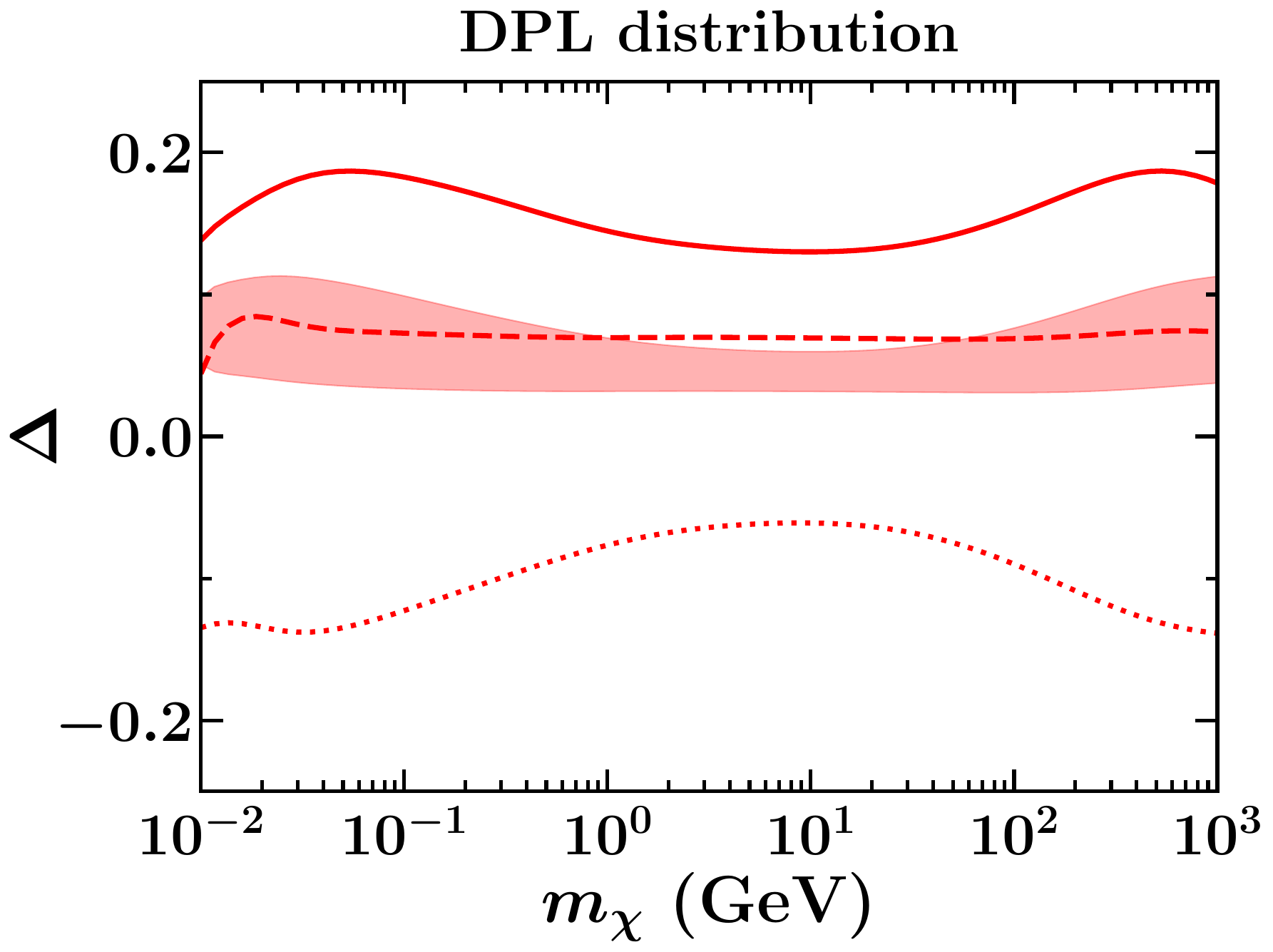} 
\caption{}
\label{sf:DPLSI}
\end{subfigure}
\begin{subfigure}{0.325\textwidth}
\centering
\includegraphics[width=1\linewidth]{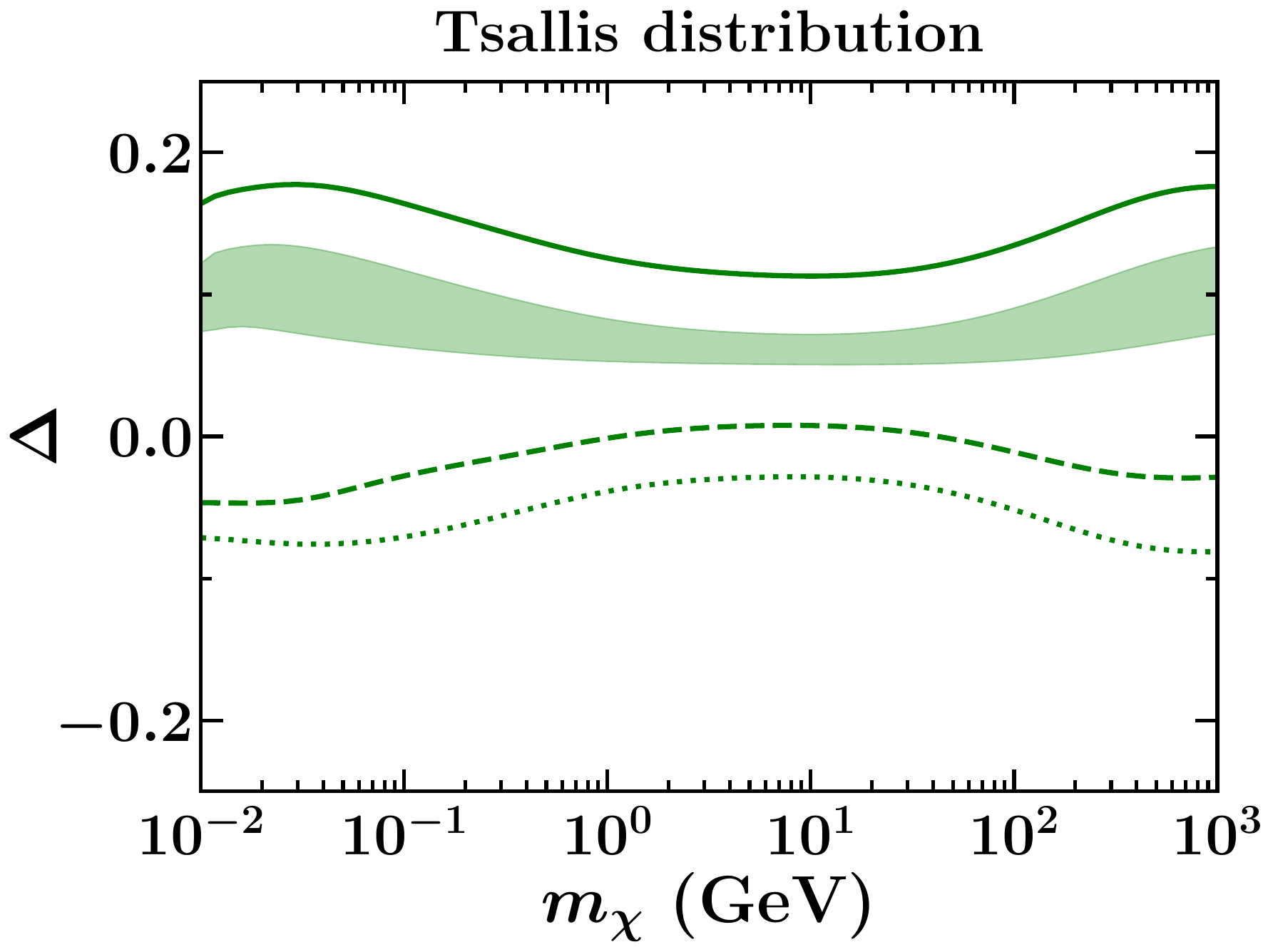} 
\caption{}
\label{sf:TsalSI}
\end{subfigure}
\begin{subfigure}{0.325\textwidth}
\centering
\hspace{0.35em} \vspace{0.55em}
\includegraphics[width=0.65\linewidth]{figs/index} 
\caption*{}
\end{subfigure}
\begin{subfigure}{0.325\textwidth}
\centering
\includegraphics[width=1\linewidth]{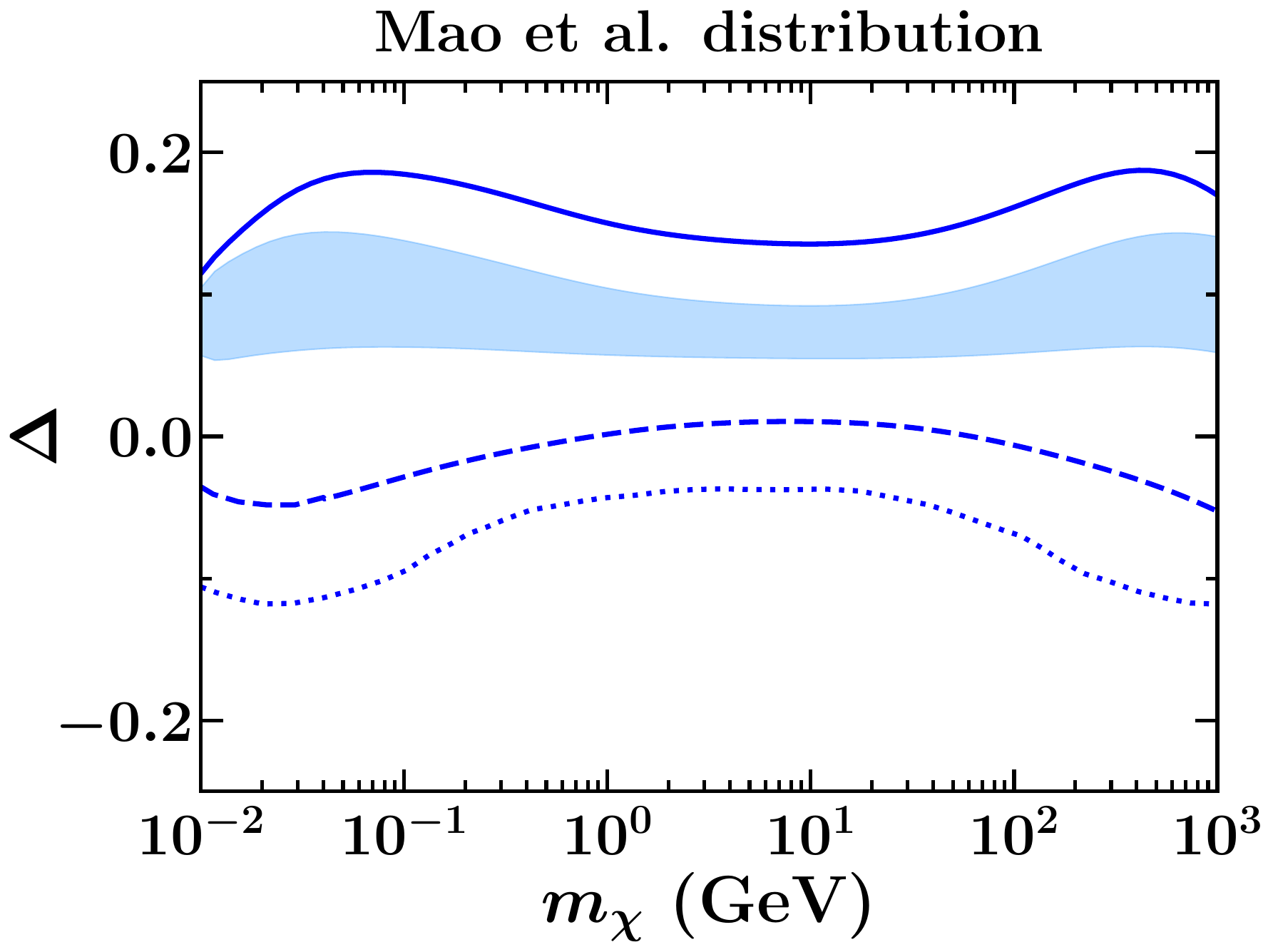} 
\caption{}
\label{sf:MaoSI}
\end{subfigure}
\caption{Same as Fig. \ref{fig:NS}, but for the Spin-Independent scenario of the Sun. The DM-nucleon scattering cross-section is fixed at $10^{-40} \, {\rm cm^2}$.}
\label{fig:SI}
\end{figure}


As presented in Figs. \ref{fig:SD} and \ref{fig:SI}, we report a change of $16\%$ and $15\%$ for the SD and SI scenarios respectively, within the range of observational uncertainties discussed in section \ref{sec:velocity_dis} in the Mao \emph{et al.} distribution. Whereas, for the FIRE-2 simulation, we find a corresponding deviation of $6\%$ and $5\%$ for the SD and SI scenarios respectively. On similar grounds the departure from the benchmark SHM for the APOSTLE and ARTEMIS simulations are at $5\%$ and $4\%$ for the SD and $23\%$ and $22\%$ for the SI scenarios respectively.

Capture of DM inside the Sun depends majorly on the solar model which accurately encompass the abundance of its various nuclei. For computing the solar capture of DM, we have used the AGSS09 model \cite{2009ARA&A..47..481A}. Implementing other motivated models may lead to alterations in the said capture rate. In order to quantify this uncertainty, we have also evaluated the capture rate using the GS98 solar model \cite{Grevesse1998}. We find a maximum uncertainty of $3\%$ and $18\%$ for the SD and SI interactions respectively. Unlike other celestial bodies discussed above, the Sun being a main sequence star, cannot be probed with dark heating signatures. However, the annihilation of the DM particles can produce neutrinos which can escape from the solar interior and reach Earth-based experiments for which the systematic uncertainties is $\mathcal{O}(10\%)$ considering the IceCube observatory \cite{IceCube:2021xzo}. The maximum uncertainty in the capture rate would lead to $19\%$ and $18\%$ variation in the SD and SI DM-nucleon scattering cross-section bound respectively, obtained by analyzing neutrino events.

\section{Comparing the results from simulations and observations}
\label{sec:comp}

Throughout the previous sections, our discussions on the capture rates motivated by cosmological simulations and astrophysical uncertainties, calls for a comparative analysis between the outcomes presented from the two. An attempt in understanding the influence of cosmological simulations on the capture rate, necessitates an a priory knowledge of their working. Both APOSTLE and FIRE-2 simulations show finer spacial and mass resolutions in comparison to most of the existing simulations, including the ARTEMIS simulation suit. A well resolved simulation involving intricate mergers, baryonic physics and accretions helps in bridging the gap between theoretical predictions and observational data. This helps us to understand as to why APOSTLE and FIRE-2 have $u_0$ and $u_{\rm esc}$ nearer to the central values determined from observations. This can be attributed to the fact as to why the capture rates estimated from APOSTLE and FIRE-2 data, closely resemble the capture rate with benchmark values of SHM. For the different distributions considered in this work including the SHM, King, DPL, Mao \emph{et al.} and Tsallis distributions, we note a maximum $\sim 24\%$ change in the capture rate for the FIRE-2 simulation for the parameters presented in Table \ref{tab:bestfit}. Whereas for the APOSTLE and ARTEMIS simulations, the number becomes $\sim 82 \%$ and $\sim 213\%$ respectively, providing a clear indication that the estimated numbers are highly correlated to the underlying dynamics governing the simulations. Incidentally, if we compute the relative change in capture rates from observational uncertainties, we arrive at a maximum $\sim 116\%$ change, relative to the benchmark SHM. Hence, these numbers computed from simulations and observation widens the prospects for detecting and constraining DM and also points at new physics beyond SHM. In Table \ref{tab:PerUnc}, we summarize the numbers obtained for SHM and non-SHM distributions for the celestial objects considered in this work. We find that for most of the halo distributions it is the astrophysical measurements which provide most significant departures from the benchmark values of SHM.

\begin{table}[t]
\centering
\resizebox{\textwidth}{!}{%
\begin{tabular}{|c|c|c|c|c|c|c|c|c|c|c|}
\hline
\multirow{2}{*}{\begin{tabular}[c]{@{}c@{}}Halo\\ Model\end{tabular}} & \multicolumn{2}{c|}{$\Delta$ (in \%) for Neutron Star} & \multicolumn{2}{c|}{$\Delta$ (in \%) for White dwarf} & \multicolumn{2}{c|}{$\Delta$ (in \%) for Brown Dwarf} & \multicolumn{2}{c|}{$\Delta$ (in \%) for  Exoplanet} & \multicolumn{2}{c|}{$\Delta$ (in \%) for the Sun} \\ \cline{2-11} 
       & N-body & Astro & N-body & Astro & N-body & Astro & N-body & Astro & N-body & Astro \\ \hline \hline
MB		& 10-17 & 2-9   & 14-17 & 1-8   & 13-18	& 2-9 	& 7-13 & 6-8    & 1-3 & 3-9 \\ \hline
King 	& 10-17 & 3-10  & 13-16 & 7-18  & 12-18	& 2-18	& 8-15 & 1-18   & 1-4 & 2-13\\ \hline
DPL	 	& 11-20 & 7-29  & 12-16 & 7-23  & 10-19	& 2-25	& 8-15 & 3-25   & 1-5 & 2-15\\ \hline
Tsallis & 9-21  & 8-49  & 10-20 & 8-48  & 1-21	& 8-49  & 8-16 & 24-49  & 1-5 & 3-15\\ \hline
Mao		& 10-22 & 8-98  & 11-22 & 8-100 & 12-23 & 8-112 & 10-18 & 37-116 & 2-6 & 3-16\\ \hline
\end{tabular}
}
\caption{Variations in the relative change in the capture rate with respect to benchmark values of MB, for the uncertainties related to astrophysical observations and the  best fit values of FIRE-2 simulation.}
\label{tab:PerUnc}
\end{table}

\section{Conclusion}
\label{sec:conclude}

WIMPs being the most popular and well explored candidate for particle DM, has provided some of the strongest constraints on DM mass and DM-SM scattering cross-section. However, the absence of a positive WIMP detection signal has motivated an avenue for the search of DM beyond ground based direct detection experiments and extending to celestial objects, stars and planets. The idea is to explore directions which can probe lower DM-nucleon cross-sections for a diverse range of DM mass. As a course of action, we systematically study the effects of observational uncertainties and cosmological simulations on the capture rate of DM within celestial objects. The accepted nomenclature is to employ a Maxwell-Boltzmann velocity distribution with typical values such as $u_0=233\, \rm km/s$ and $u_{\rm esc}=528 \, \rm km/s$, in order to detect, explore and constrain the properties of particulate DM. We report digressions in capture rate from this standard choice due to the introduction of observational uncertainties and non-standard, isotropic, distributions which are motivated from well resolved and sophisticated cosmological simulations. Additionally, we probe different DM-nucleon scattering cross-sections inside a diverse class of celestial objects. Apart from the astrophysical uncertainties which imparts the main essence to this work, other model parameters like radius, mass and equation of state of the celestial objects can also introduce variations in the capture rates. We have provided some back of the envelope estimates on the change in capture rate due to these additional effects wherever possible. We have found them to yield $\mathcal{O}(10\%)$ changes in the capture rate, which is less in comparison to the figures derived from variations in velocity distributions and its associated parameters. Although a detailed approach accounting for all possible uncertainties would make such a study more robust, it would require a dedicated introspection and hence remains beyond the mandate of this work.

The rate at which ambient DM losses its energy in order to thermalize and eventually get captured and annihilated inside celestial objects, is conjectured to be sensitive to the population of DM particles, particularly at the low velocity tail of its distribution. This in turn depends on the Sun's circular velocity $u_0$, giving the velocity distribution its characteristic spread. Remaining within the chassis of SHM and astrophysical errors, we report a maximum $\sim 10 \%$ change in capture rate from astrophysical uncertainties. Whereas, well a resolved and detailed cosmological simulations like FIRE-2 can invoke a maximum  $\sim 24 \%$ change in the capture rate. We expect this number to go up if the resolution of the simulations and the processes involving structure formation are not taken into account with significant accuracy. 

As we move beyond SHM, empirical, isotropic, non-SHM distributions provide larger variations in the relative capture rate with respect to the benchmark SHM. All non-MB distributions like the King, DPL, Mao \emph{et al.} and Tallis fall smoothly near the velocity tails, predicting less number of particles in those regions. Therefore, for the same set of astrophysical parameters, the non-MB models seem to predict a reduction in the rate at which DM gets capture as compared to their MB counterpart. In case of the non-standard distributions we report a  $\sim 116 \%$ change in the capture rate as a result of the uncertainties related to relevant observations. However, this number can be as large as $\sim 213 \%$ for the ARTEMIS simulation. The results obtained from our analysis hints at significant reinterpretation of the conclusions concerning the present day and upcoming indirect searches of DM.

\paragraph*{Acknowledgments\,:} 

We thank Somnath Bharadwaj, Tirtha Sankar Ray and Tarak Nath Maity for the helpful discussions. We would also like to thank Sandra Robles for the valuable comments. DB acknowledges MHRD, Government of India for fellowship. SS acknowledges the University Grants Commission (UGC) of the Government of India for providing financial assistance through Senior Research Fellowship (SRF), with reference ID: 522157.

\appendix


\section{Simulations employed}
\label{sec:app1}

\begin{enumerate}

\item \textbf{APOSTLE}:

We have used the values for the halo A1, identified as a MW-like halo from the APOSTLE (A Project Of Simulating The Local Environment) simulation suit \cite{Sawala:2015cdf}, the values of which have been adapted from \cite{Maity:2020wic}. The selection criterion is based on the conditions that the total halo mass ($M_{200}$) in the range $5 \times 10^{11} < M_{200}/M_{\odot} < 2 \times 10^{13}$, having rotation curves similar to the observed MW like rotation curves, stellar mass ($M_{\star}$) in the observed mass range $4.5 \times 10^{10} < M_{\star}/M_{\odot} < 8.3 \times 10^{10}$, having a substantial stellar disc \cite{Bozorgnia:2016ogo}. Apostle uses three different levels of mass resolution for the primordial gas (DM) particles, namely $1.0(5.0) \times 10^4 \, M_{\odot}$, $1.2(5.9) \times 10^5  \, M_{\odot} $ and $1.5(7.5) \times 10^6  \, M_{\odot}$ respectively. Maximum gravitational softening lengths of $134$ pc, $307$ pc and $711$ pc.

\item \textbf{ARTEMIS}:

The ARTEMIS (Assembly of high-ResoluTion Eagle-simulations of MIlky Way-type galaxieS) simulation suit \cite{Font_2020} studied the evolution of 42 MW-like haloes, in the mass range of $8 \times 10^{11} < M_{200}/M_{\odot} < 2\times 10^{12}$. For this work we have used the parameters stated in \cite{Maity:2020wic}, which is for the median DM distribution of the 42 MW-like galaxies discussed in \cite{Poole-McKenzie:2020dbo}. ARTEMIS is one of the largest suit of well resolved cosmological simulation run till date, with both baryons and DM particles of masses $2.2 \times 10^4 \,\rm M_{\odot }h^{-1}$ and $1.2 \times 10^5 \, \rm M_{\odot}h^{-1}$, respectively.

\item \textbf{FIRE-2}:

The FIRE (Feedback In Realistic Environments)-2 simulations are run with the GIZMO code \cite{Hopkins:2014qka} using the mesh free finite-mass (MFM) Lagrangian Godunov method for hydrodynamics, while gravitational interactions are solved using the Tree-PM code GADGET-3 \cite{Springel:2005mi}. We use the main halo from the \texttt{m12f} simulation \cite{Garrison_Kimmel_2017}, which experiences more mergers at late cosmological times. The $m12f$ assume a $\Lambda$CDM cosmology with $\Omega_{\Lambda} = 0.728\,, \Omega_m = 0.272 \,, \Omega_b = 0.0455\,, h = 0.702\,, \sigma = 0.807$, and $n_s = 0.961$. The  host halo in $m12f$ has a virial mass of $M_{200} = 1.7 \times 10^{12}\, M_{\odot}$, virial radius $ R_{200} = 380 \,\rm kpc$ and DM particle number $7.44 \times 10^7$ \cite{Necib:2018igl}. The mass of baryonic particles is approximately $7070 \,M_{\odot}$ and $5000 \,M_{\odot}$ for the stars. The Plummer equivalent gravitational softening is $4$ pc for stars and  $1$ pc for gas particles. DM particles in the zoom-in region have mass resolution of $3.5 \times 10^4 \,M_{\odot}$ and a softening length of $40$ pc. The total stellar mass of the Galactic disk of the \texttt{m12f} host halo is $6.9 \times 10^{10}  \,M_{\odot}$.

\end{enumerate}


\section{Optically thin and geometric limit}
\label{sec:app2}

\begin{figure}[t]
\centering
\begin{subfigure}{0.325\textwidth}
\centering
\includegraphics[width=1\linewidth]{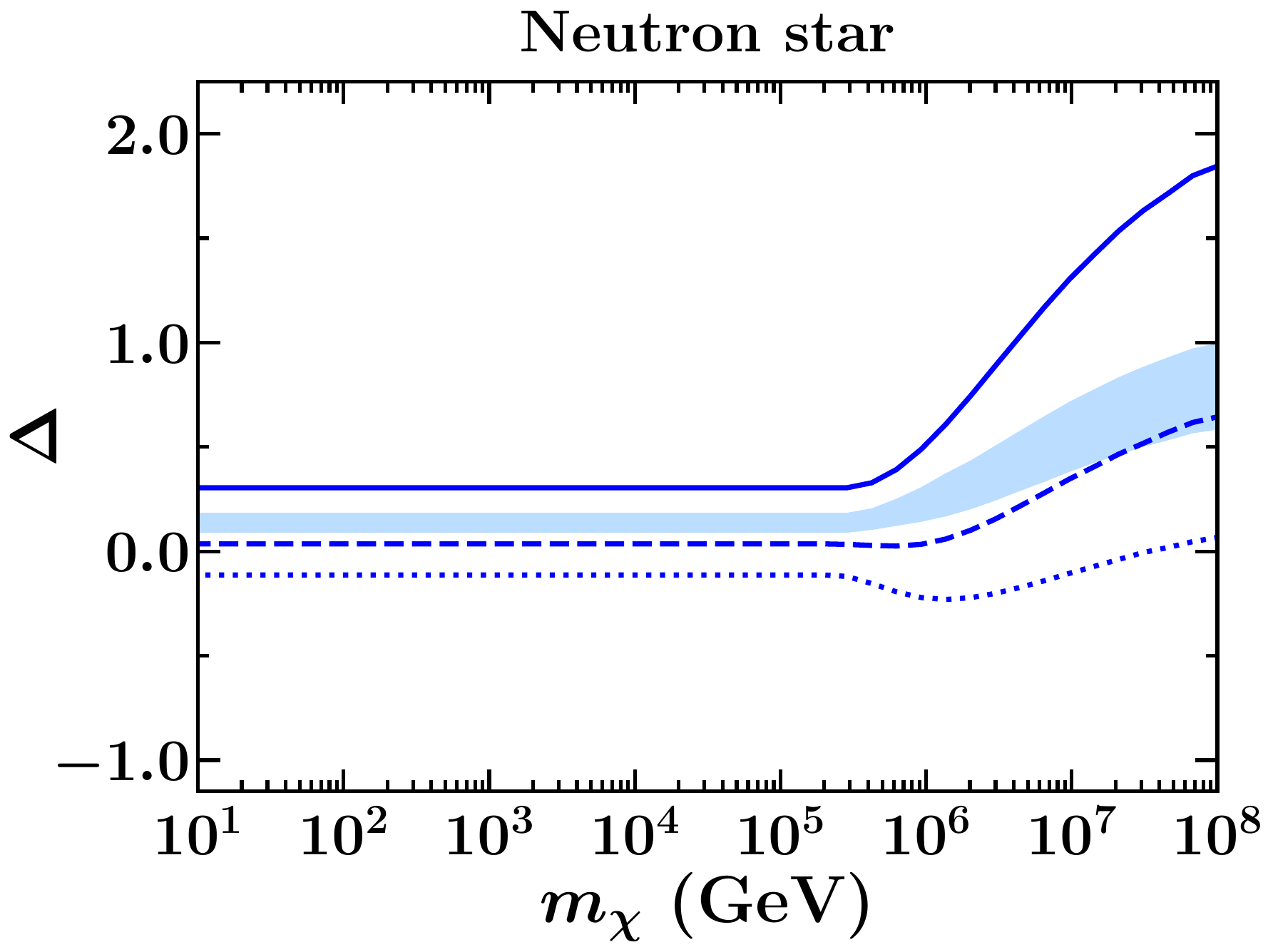} 
\caption{}
\label{sf:MBSI}
\end{subfigure}
\begin{subfigure}{0.325\textwidth}
\centering
\hspace{0.35em} \vspace{0.55em}
\includegraphics[width=0.65\linewidth]{figs/index} 
\caption*{}
\end{subfigure}
\begin{subfigure}{0.325\textwidth}
\centering
\includegraphics[width=1\linewidth]{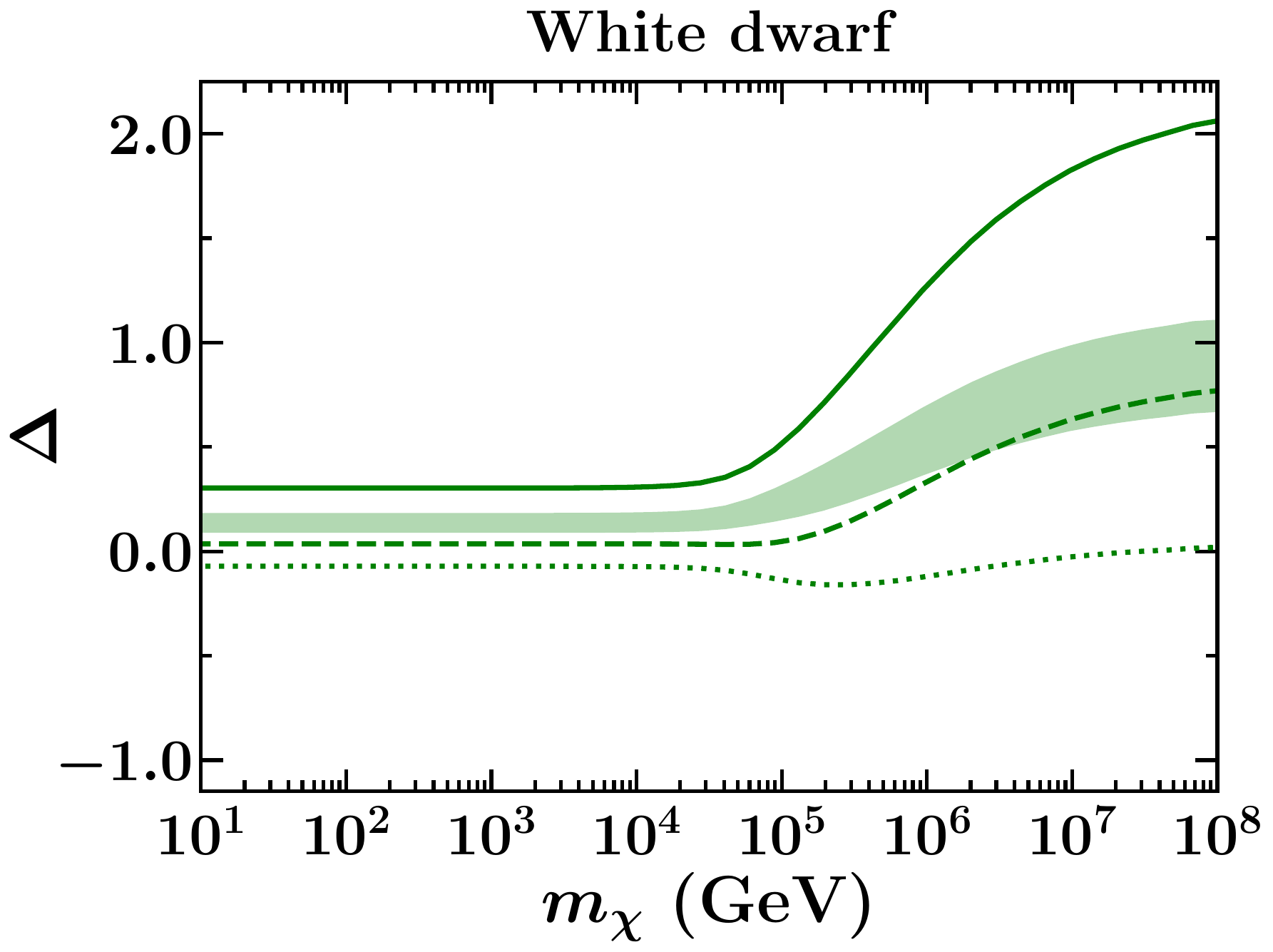} 
\caption{}
\label{sf:DPLSI}
\end{subfigure}
\begin{subfigure}{0.325\textwidth}
\centering
\includegraphics[width=1\linewidth]{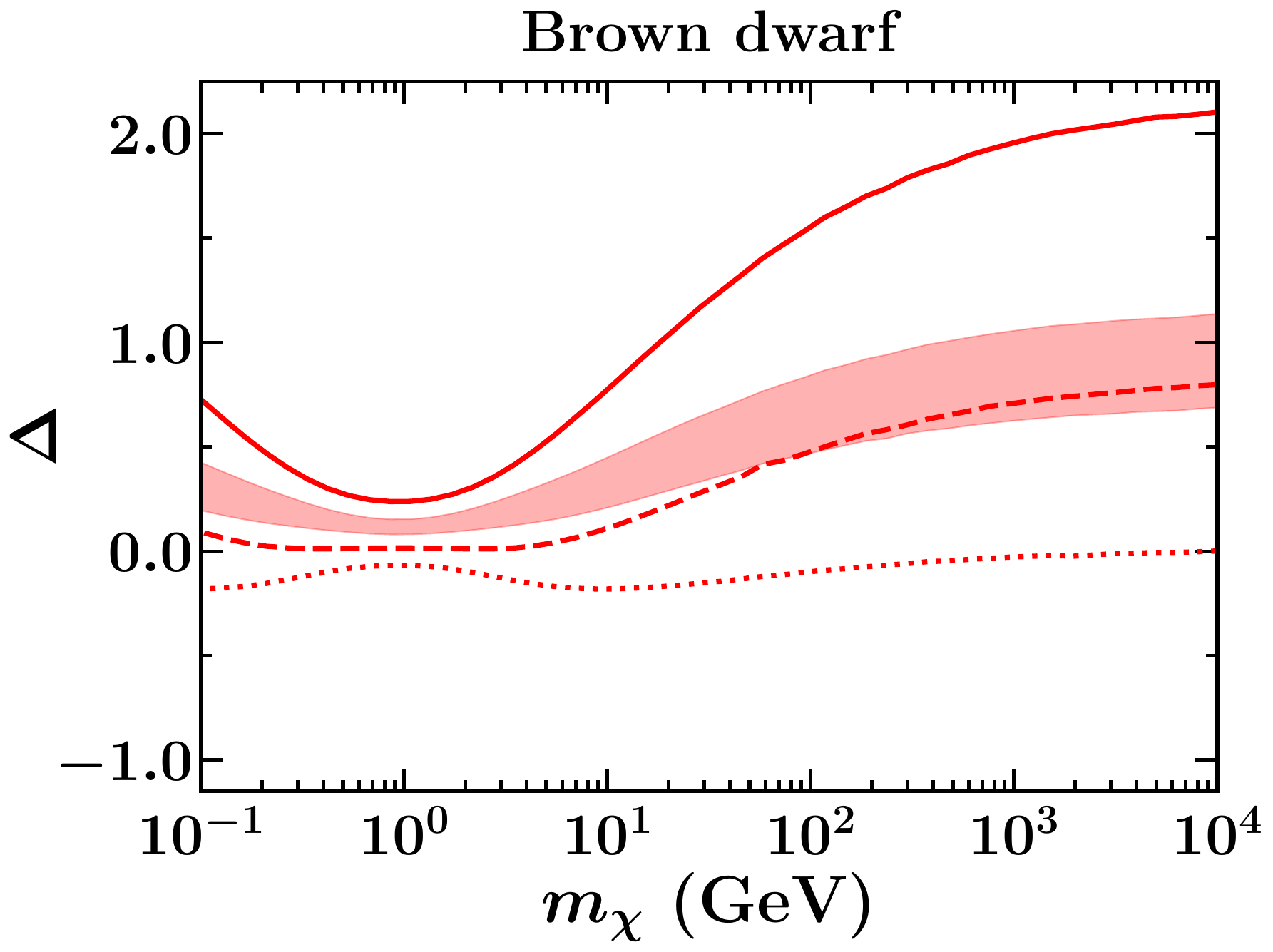} 
\caption{}
\label{sf:TsalSI}
\end{subfigure}
\begin{subfigure}{0.325\textwidth}
\centering
\includegraphics[width=1\linewidth]{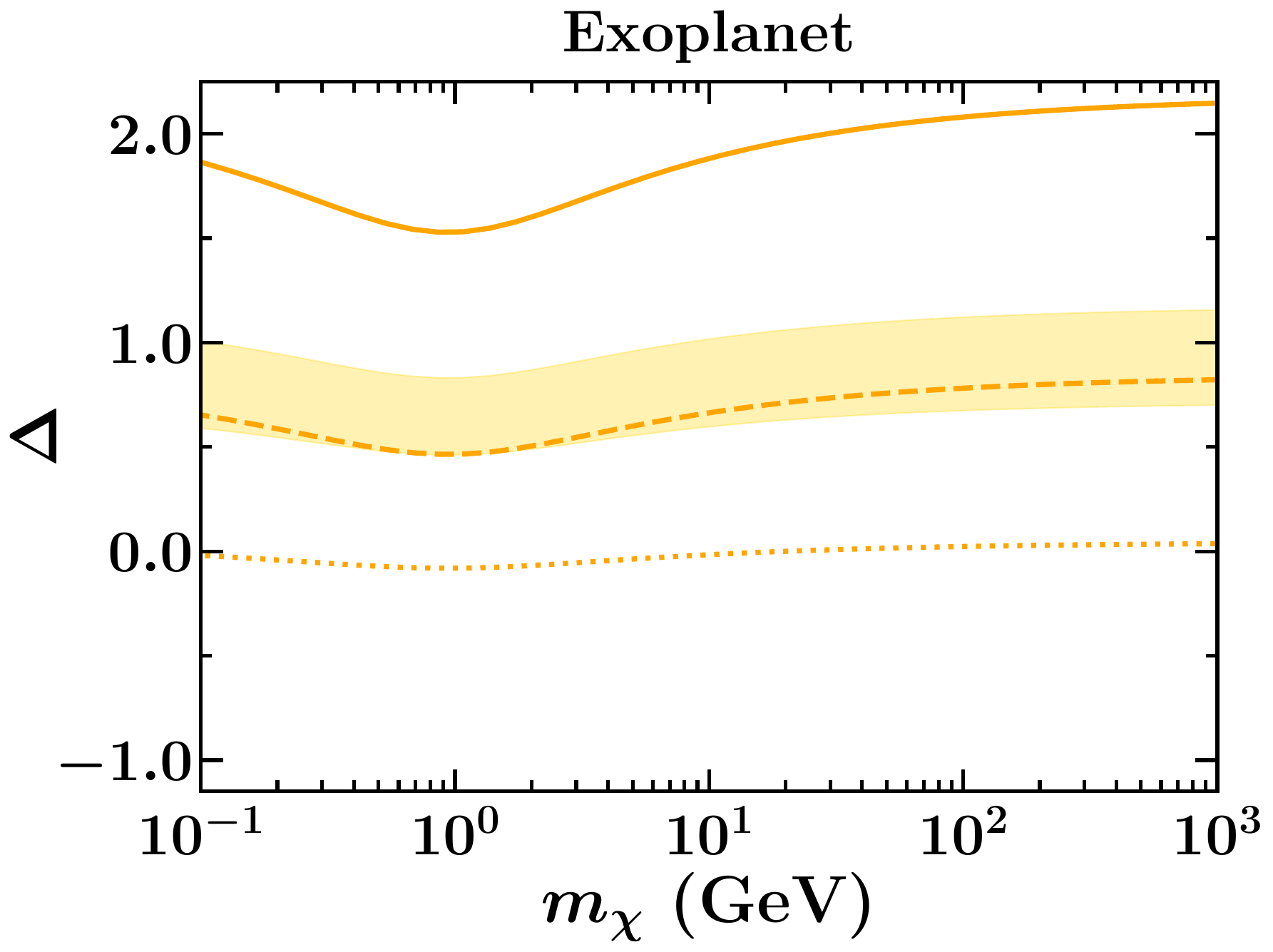} 
\caption{}
\label{sf:KingSI}
\end{subfigure}
\caption{The relative change in the DM capture rate with respect to the MB benchmark values as computed from Eq. \eqref{eq:capt-effective}, for the neutron star, white dwarf, brown dwarf and exoplanet depicted by blue, green, red and yellow colored plots respectively. The shaded band in each plot refers to the variation in the relative capture rate by taking into account the astrophysical uncertainty in the estimates of $u_0$ and $u_{\rm esc}$. The solid, dashed and dotted lines show the variations for the ARTEMIS, APOSTLE and FIRE-2 simulations respectively. We have considered the mass and radius of celestial bodies to be same as before and the DM-nucleon scattering cross-section for NS, WD, BD and exoplanet is kept at $10^{-47}$, $10^{-40}$, $10^{-38}$ and $10^{-36} \, {\rm cm^2}$ respectively.}
\label{fig:app1}
\end{figure}

In this section, we discuss and present our findings on the relative change in capture rate inside the celestial bodies, described in section \ref{sec:cap_all} for the DM-nucleon scattering cross-section below and above the saturation cross-section. Following Eq. \eqref{eq:caprate}, we calculate the capture rates below the saturation cross-section, i.e in the optically thin limit. While the mass and radius of the celestial bodies remain unchanged, the DM-nucleon scattering cross-section is reduced to $10^{-47} \, {\rm cm^2}$ for NS, $10^{-40} \, {\rm cm^2}$ for WD, $10^{-38} \, {\rm cm^2}$ for BD and $10^{-36} \, {\rm cm^2}$ for the exoplanet. As the relative change is more prominent for the Mao \emph{et al.} distribution, we have shown the variations in $\Delta$, with DM mass, for different celestial bodies in figure \ref{fig:app1}. The blue, green, magenta and yellow colored plots are for the NS, WD, BD and exoplanet respectively. The solid, dashed and dotted lines depict the relative changes for ARTEMIS, APOSTLE and FIRE-2 respectively, while the shaded regions show the variations that come from the uncertainty in the astrophysical observations.

For a DM-nucleon scattering cross-section much greater than the saturation cross-section, the coupling between DM and nucleon becomes so strong that the capture rate can be approximated to be in the geometric limit of the celestial object. This implies that all the DM particles which enter the celestial body get captured. The geometric capture rate is given by \cite{Garani:2018kkd,Lopes:2020dau}
\begin{equation}
\label{eq:caprate_geo}
C_{\rm geo} = \pi \, R^2 \, \left( \frac{\rho_{\chi}}{m_{\chi}} \right) \int_{0}^{u_{\rm esc}} \dfrac{f(u_{\chi}) \, du_{\chi}}{u_{\chi}} \, \left( u_{\chi}^2+v_{\rm esc}^2 \right),
\end{equation}
where the symbols carry the meanings stated earlier. It is evident from Eq. \eqref{eq:caprate_geo} that the relative change in the geometric capture rate is independent of the mass. In Table \ref{tab:geolimit}, we have cataloged the relative change in the geometric capture rate for different celestial bodies, considering the Mao \emph{et al.} distribution.

\begin{table}[h]
\centering
\begin{tabular}{|c|c|c|c|c|}
\hline
\multirow{2}{*}{\begin{tabular}[c]{@{}c@{}} Celestial body \end{tabular}} & 
\multirow{2}{*}{\begin{tabular}[c]{@{}c@{}} $\Delta$ (in \%) for astrophysical\\ uncertainty \end{tabular}} & 
\multicolumn{3}{c|}{$\Delta$ (in \%) for Simulation}  \\ \cline{3-5} 
             &   & APOSTLE & ARTEMIS & FIRE-2 \\ \hline \hline
Neutron star & 18.45 & 3.56 & 30.44 & 7.19 \\
\hline
White dwarf & 18.39 & 3.56 & 30.34 & 7.16 \\
\hline
Brown dwarf & 11.53 & 3.11 & 19.67 & 3.92 \\
\hline
Exoplanet & 12.54 & 1.53 & 17.78 & 7.42 \\
\hline
\end{tabular}
\caption{Relative changes in the capture rate $\Delta$, in the geometric limit for Mao \emph{et al.} distribution.}
\label{tab:geolimit}
\end{table}

\vspace*{0.4 cm}

\section{Parametric dependence on capture rate}
\label{sec:app3}

\begin{figure}[h]
\centering
\begin{subfigure}{0.465\textwidth}
\centering
\includegraphics[width=1\linewidth]{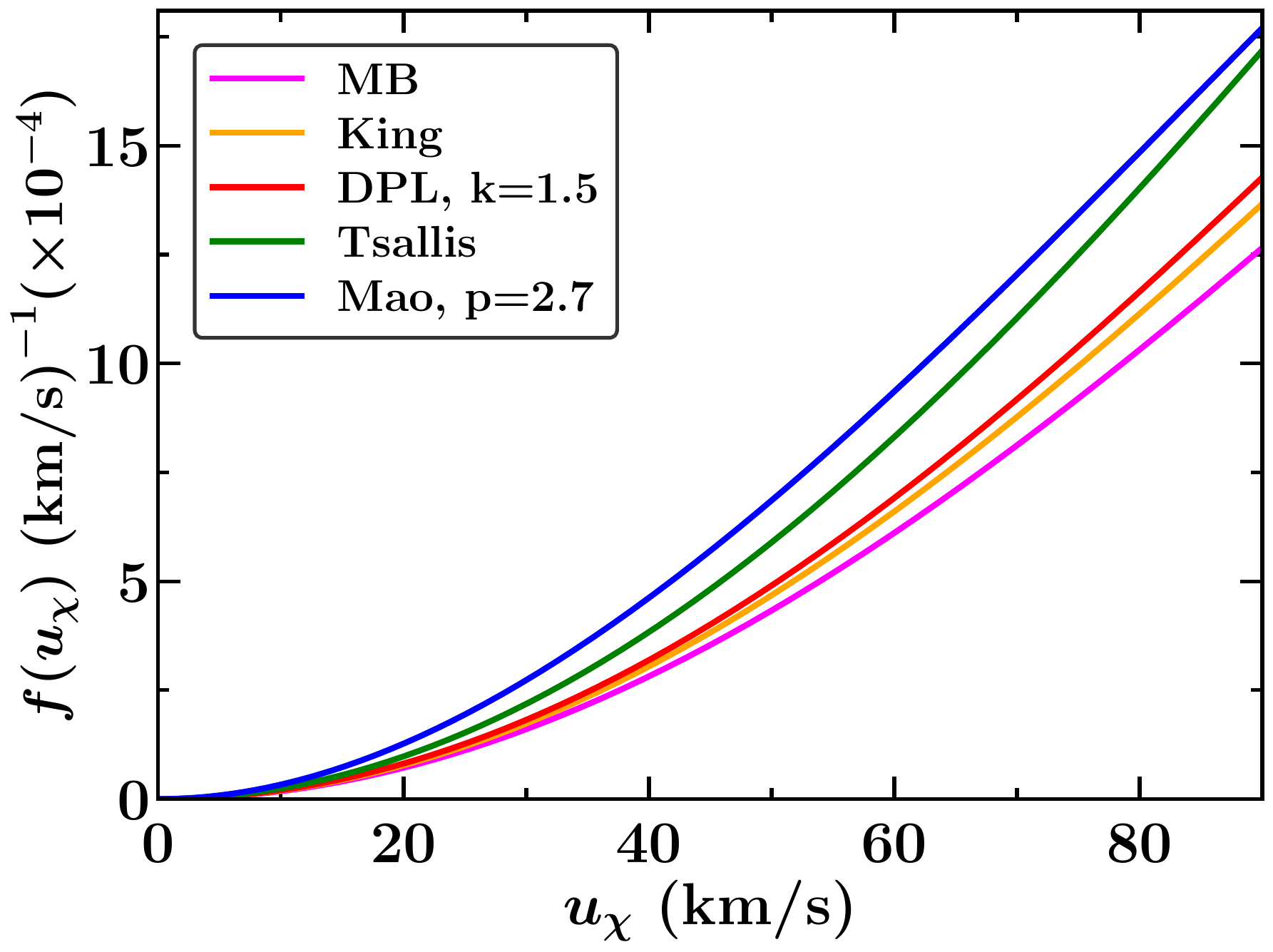} 
\caption{}
\label{sf:comp}
\end{subfigure}
\hspace*{0.15 cm}
\begin{subfigure}{0.465\textwidth}
\centering
\includegraphics[width=1\linewidth]{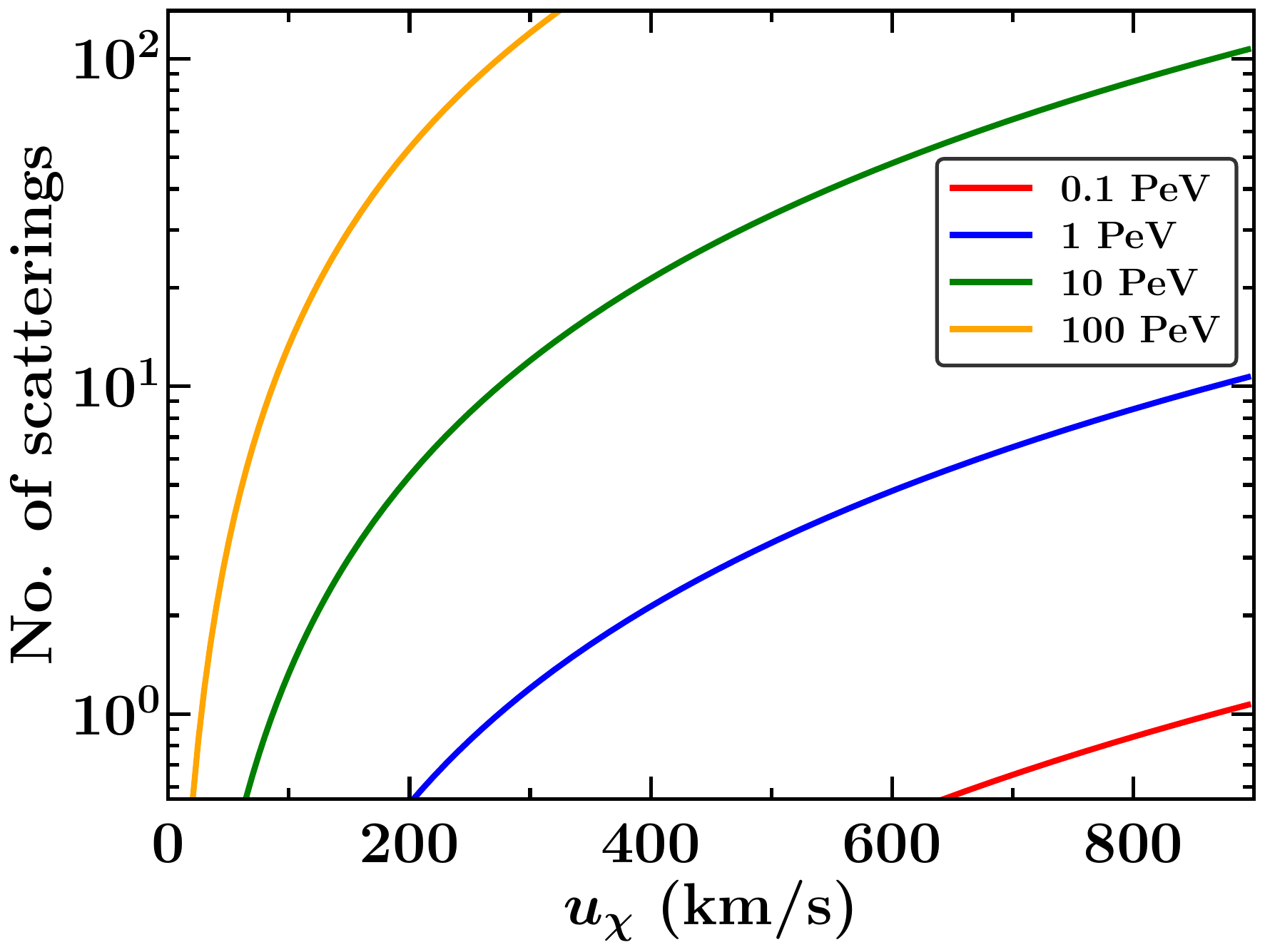}
\caption{}
\label{sf:scat}
\end{subfigure}
\caption{\textit{Left panel}: Variation of the normalized DM velocity distribution as a function of DM velocity for the benchmark choice of parameters. \textit{Right panel}: Contours showing the required number of scatterings as a function of DM velocity at DM masses of 0.1 PeV, 1 PeV, 10 PeV and 100 PeV, shown by the red, blue, green and yellow contours respectively.}
\label{fig:app2}
\end{figure}

In this section, we discuss how the astrophysical parameters appearing in equation \eqref{eq:caprate}, impact the rate at which massive DM particles gets captured. As discussed in section \ref{sec:velocity_dis} the effect of $u_0$ and $u_{\rm esc}$ on the capture rate can be understood from figure \ref{sf:comp}. For the same set of astrophysical parameters, the area under the curve can be seen to maximum for the Mao \emph{et al.} distribution, followed by the Tsallis DPL, King and MB distributions respectively. This gets reflected when we look into the deviations produced by the non-MB distributions with respect to the benchmark values of MB. In figure \ref{sf:scat} we plot the required number of scatters required for a DM particle of mass $m_{\chi}$ to get entrapped by the gravitational potential of the NS in consideration, traveling with an initial velocity $u_{\chi}$. The criteria is given by,
\begin{equation}
\label{escapevel}
g_{N}(w)=\Theta \left(v_{\rm esc}(1-\frac{1}{2}\beta_{+})^{-\frac{N}{2}}-w\right)
\end{equation}

Here the symbols carry their usual meaning and $\beta_{+}=\frac{4m_t m_{\chi}}{(m_t + m_{\chi})^2}$, $m_t$ being the mass of the target nuclei. Therefore, a DM particle of mass $100$ PeV, with an incumbent velocity $\sim 90 \, \rm km/s $ would require atleast $\sim 10 $ scatters before its velocity becomes less than the escape velocity of the NS.


\bibliographystyle{JHEP}
\bibliography{uncertaincap.bib}

\end{document}